\documentclass[superscriptaddress,aps,preprintnumbers,showpacs,prd,nofootinbib,preprint]{revtex4-1}

\usepackage{graphicx} 
\usepackage{amsmath,amssymb,amsthm,slashed,bm}
\usepackage[colorlinks,allcolors=blue]{hyperref} 

\begin{document}

%%%%%%%%%%%%%%%%%%%%%%%%%%%%%%%%%%%%%%%%%%%

\preprint{TU-1315, RIKEN-iTHEMS-Report-26}

\title{ 
Axion Isocurvature Perturbations Survive the Scaling Evolution of Axion Domain Walls
}

\author{
Naoya Kitajima
}
\affiliation{Frontier Research Institute for Interdisciplinary Sciences, Tohoku University, Sendai, Miyagi 980-8578, Japan}
\affiliation{Department of Physics, Tohoku University, 
Sendai, Miyagi 980-8578, Japan} 
\author{
Junseok Lee
}
\affiliation{Department of Physics, Tohoku University, 
Sendai, Miyagi 980-8578, Japan}
\author{
Kai Murai
}
\affiliation{RIKEN Center for Interdisciplinary Theoretical and Mathematical Sciences (iTHEMS), RIKEN, Wako 351-0198, Japan}
\affiliation{Department of Physics, Tohoku University, 
Sendai, Miyagi 980-8578, Japan} 
\author{
Fuminobu Takahashi
}
\affiliation{Department of Physics, Tohoku University, 
Sendai, Miyagi 980-8578, Japan} 
\affiliation{Kavli IPMU (WPI), UTIAS, University of Tokyo, Kashiwa 277-8583, Japan}
\author{
Wen Yin
}
\affiliation{Department of Physics, Tokyo Metropolitan University,
Minami-Osawa, Hachioji-shi, Tokyo 192-0397, Japan}

\begin{abstract}
We revisit the evolution of axion domain walls seeded by inflationary fluctuations.
In our previous work, we showed that such domain-wall networks retain superhorizon correlations even after entering the scaling regime.
We extend our previous analysis to the case with large initial fluctuations, where many minima of the axion potential are already populated when the axion starts to oscillate.
Although the conventional misalignment contribution can have suppressed long-wavelength isocurvature perturbations when many vacua are averaged over, axions produced by domain-wall collapse provide an additional contribution that can dominate  when the walls enter the scaling regime before annihilation.
In particular, the biased vacuum energy released during wall annihilation inherits the superhorizon correlations of the inflationary fluctuations and transfers them to the axion energy density.
We find that sizable isocurvature perturbations can therefore survive even after the walls annihilate.
We also discuss generic isocurvature constraints on dark matter produced by domain-wall collapse.
\end{abstract}

\maketitle
\flushbottom

\vspace{1cm}

%%%%%%%%%%%%%%%%%%%%%%%%%%%%%%%%%%%%%%
\section{Introduction}
\label{sec: intro}
%%%%%%%%%%%%%%%%%%%%%%%%%%%%%%%%%%%%%%

In the early universe, many symmetries are expected to have been restored and then spontaneously broken as the universe expanded and cooled. When such symmetry breaking leads to the formation of topological defects, they can be long-lived because the topology of the vacuum manifold prevents them from unwinding~\cite{Kibble:1976sj,Kibble:1980mv,Vilenkin:1984ib}. They therefore play an important role in cosmology as messengers of high-energy physics in the early universe, for instance, through the production of gravitational waves and dark matter.

A key concept in understanding the cosmological evolution of topological defects is the scaling behavior. In this paper, we focus on domain walls, for which scaling means that the network has a characteristic separation and curvature radius of order the Hubble radius. Numerical studies have shown that domain-wall networks approach such a scaling regime for many conventional initial conditions~\cite{Press:1989yh,Hindmarsh:1996xv,Garagounis:2002kt,Oliveira:2004he,Avelino:2005kn,Avelino:2008ve,Leite:2011sc,Leite:2012vn,Martins:2016ois}. Similar scaling behavior is also familiar for other topological defects, such as cosmic strings and monopoles.

There is a standard lore associated with scaling behavior: the scaling regime is often regarded as a local attractor, and defect networks are therefore expected to lose memory of their initial conditions. This expectation is indeed well supported when defects are generated through the usual Kibble mechanism~\cite{Kibble:1976sj,Kibble:1980mv}, where the initial fluctuations are thermal or effectively white-noise-like and are dominated by scales much smaller than the Hubble radius. We emphasize, however, that this statement does not apply to arbitrary initial conditions. 

Another physically important possibility is that the initial fluctuations are generated during inflation. In our previous work, taking domain walls as an example, we showed that field correlations initially present on superhorizon scales remain imprinted in the subsequent evolution of the defect network~\cite{Gonzalez:2022mcx,Kitajima:2023kzu}. This is in fact what one should expect from causality: local defect dynamics cannot erase correlations between regions that have never been in causal contact. We stress that there is no contradiction between scaling and the persistence of such correlations; even if a network has a characteristic separation and curvature radius of order the Hubble radius, it can still carry correlations on superhorizon scales. The widespread impression that scaling necessarily implies loss of memory appears to originate from conventional applications of the Kibble mechanism to thermal or white-noise initial conditions.\footnote{Note that the Kibble mechanism~\cite{Kibble:1976sj} was proposed before inflationary cosmology was developed~\cite{Guth:1980zm,Starobinsky:1980te,Sato:1981qmu}.} One of the main purposes of this paper is to clarify that entering the scaling regime does not necessarily imply the erasure of primordial superhorizon correlations.

For simplicity, in this paper, we focus on the evolution of domain walls. Domain walls require multiple degenerate, or nearly degenerate, minima in the potential. Axions provide a natural setting for such a situation: their shift symmetry is explicitly broken by a periodic potential, and, depending on the domain-wall number or the structure of the potential, multiple vacua can appear. Axions are also known as a good candidate for dark matter~\cite{Peccei:1977hh,Peccei:1977ur,Weinberg:1977ma,Wilczek:1977pj,Preskill:1982cy,Abbott:1982af,Dine:1982ah}, and
axion topological defects have been extensively studied in connection with axion dark matter and gravitational waves production~\cite{Sikivie:1982qv,Nagasawa:1994qu,Larsson:1996sp,Chang:1998tb,Hiramatsu:2010yn,Hiramatsu:2012sc,Hiramatsu:2012gg,Kawasaki:2014sqa,Gorghetto:2018myk,Buschmann:2021sdq,Chang:2023rll,Gorghetto:2023vqu,Saikawa:2024bta}. Recently, the axion domain walls have attracted attention from the viewpoint of multiple axions~\cite{Daido:2015gqa,Daido:2015bva,Higaki:2016jjh,Kitajima:2023cek,Lee:2024xjb,Lee:2024toz,Mariotti:2024eoh,Lee:2025zpn,Lee:2026umy,FernandezNavarro:2026cyu,Banks:2026yhy}.

In this work, we consider axion domain walls that are not accompanied by cosmic strings. This is distinct from the usual post-inflationary scenario, in which axion strings and domain walls form as a connected defect network. 
String-free axion domain walls can arise through several mechanisms. A simple possibility is that inflationary fluctuations of a light axion populate different branches of the periodic potential before the axion starts to oscillate~\cite{Linde:1990yj,Lyth:1991ub,Nagasawa:1991zr,Higaki:2016yqk,Takahashi:2020tqv,Gonzalez:2022mcx}. In multi-axion systems, domain walls may also form after nonadiabatic amplification of field fluctuations around a level crossing, a phenomenon known as axion roulette~\cite{Daido:2015cba,Daido:2015bva}. More generally, related wall configurations can arise from large inhomogeneous axion or ALP field configurations with nonzero momenta~\cite{Narita:2025jeg,Miyazaki:2025tvq}, and such inhomogeneities can be enhanced when the decay constant is dynamical~\cite{Kobayashi:2016qld}. String-free axion domain walls are also relevant in applications to cosmic birefringence~\cite{Takahashi:2020tqv,Kitajima:2022jzz,Gonzalez:2022mcx,Ferreira:2023jbu,Lee:2025yvn}.

In our previous work, we studied a setup in which inflationary fluctuations around the top of the potential lead to domain-wall formation, and showed that the associated superhorizon correlations survive even after the domain-wall network reaches the scaling regime~\cite{Gonzalez:2022mcx}.
Here we extend our previous analysis to axion domain walls with large initial inflationary fluctuations, so that multiple minima are populated from the outset. Similar setups have been considered previously~\cite{Linde:1990yj,Lyth:1991ub,Takahashi:2020tqv,Gonzalez:2022mcx,Kitajima:2023kzu} and have recently been revisited in Ref.~\cite{Karananas:2025uhy}. In line with the causal picture established in our previous work, we show that axion isocurvature perturbations are not erased by domain-wall evolution. More concretely, even after a potential bias makes the axion domain walls collapse, the resulting axion abundance retains fluctuations on superhorizon scales. Thus, even if the isocurvature perturbation in the axion energy density produced by the misalignment mechanism is suppressed when many vacua are populated~\cite{Kofman:1985zx,Kofman:1986wm,Linde:1990yj}, a sizable isocurvature perturbation is generated by axions produced by domain-wall decay. This poses a serious challenge to the recently proposed mechanism for evading axion isocurvature constraints in Ref.~\cite{Karananas:2025uhy}.

The rest of this paper is organized as follows. In Sec.~\ref{sec: setup}, we introduce the axion potential and the bias term, and classify the domain-wall evolution into the single- and multiple-bias-period regimes. In Sec.~\ref{sec:fate-of-DWs}, we discuss the fate of the domain-wall networks in these regimes, including the bias-driven collapse, the transfer of primordial superhorizon correlations to the axion energy density, and the formation and delayed collapse of composite domain walls. In Sec.~\ref{sec: lattice}, we present numerical lattice simulations of domain walls seeded by scale-invariant inflationary fluctuations and study the resulting axion density perturbations. In Sec.~\ref{sec:isocurvature-formation}, we derive generic isocurvature constraints on dark matter produced by domain-wall collapse and discuss the prospects for probing spectra with different large-scale tilts. Finally, Sec.~\ref{sec: conclusion} is devoted to discussion and conclusions.

%%%%%%%%%%%%%%%%%%%%%%%%%%%%%%%%%%%%%%
\section{Setup}
\label{sec: setup}
%%%%%%%%%%%%%%%%%%%%%%%%%%%%%%%%%%%%%%

We consider an axion field $\phi$ with decay constant $f_\phi$,
whose potential is given by
\begin{align}
V(\phi)
=
m_\phi^2 f_\phi^2
\left[
1-\cos\left(\frac{\phi}{f_\phi}\right)
\right],
\end{align}
where $m_\phi$ is the axion mass.
The key assumption in this paper is that the axion is not embedded
in a complex scalar field, but is instead treated as a real scalar
field with a fundamental shift symmetry.
Thus, its configuration space does not contain vortex configurations.
This setup is analogous to that of the QCD axion when the
Peccei--Quinn (PQ) symmetry is broken before or during inflation and is
never restored afterward.
If the axion is effectively massless during inflation, it acquires
nearly scale-invariant fluctuations with an amplitude of order
$H_{\rm inf}/2\pi$, where $H_{\rm inf}$ is the Hubble parameter
during inflation.
These fluctuations generally induce isocurvature perturbations in
axion dark matter~\cite{Turner:1990uz,Kobayashi:2013nva}, which are
constrained by cosmic microwave background (CMB)
observations~\cite{Planck:2018jri}.

We define the power spectrum of the axion fluctuation by
\begin{align}
 \left\langle
 \delta\phi({\bf k})\delta\phi({\bf k}')
 \right\rangle
 =
 (2\pi)^3\delta^{(3)}({\bf k}+{\bf k}')P_\phi(k),
\end{align}
and the corresponding dimensionless power spectrum by
\begin{align}
 \Delta_\phi^2(k)
 \equiv
 \frac{k^3}{2\pi^2}P_\phi(k).
\end{align}
Here, $\bf{k}$ is the comoving wave number vector, and $k$ is its modulus. 
For nearly scale-invariant inflationary fluctuations,
\begin{align}
 P_\phi(k)
 \simeq
 \frac{2\pi^2}{k^3}
 \left(\frac{H_{\rm inf}}{2\pi}\right)^2,
 \qquad 
 \Delta_\phi^2(k)
 \simeq
 \left(\frac{H_{\rm inf}}{2\pi}\right)^2.
\end{align}
For later use, we also define the corresponding two-dimensional spectrum such that it gives the same field variance per logarithmic interval in momentum:
\begin{align}
P_\phi^{(2{\rm D})}(k)
\simeq
\frac{2\pi}{k^2}
\left(\frac{H_{\rm inf}}{2\pi}\right)^2,
\qquad
\Delta_{\phi,2{\rm D}}^2(k)
\equiv
\frac{k^2}{2\pi}P_\phi^{(2{\rm D})}(k)
\simeq
\left(\frac{H_{\rm inf}}{2\pi}\right)^2.
\end{align}

When the Hubble parameter during inflation is much larger than the axion decay constant, $H_{\rm inf} \gg 2\pi f_\phi$, inflationary fluctuations of the axion field are sufficiently large that different patches of the universe can later settle into different minima of the axion potential.%
\footnote{
If the axion is embedded in a complex scalar field, this situation corresponds to the case where the domain-wall number $N_{\rm DW}$ is so large that $f_\phi = F_\phi / N_{\rm DW}$ is much smaller than $H_\mathrm{inf}$ while $F_\phi$ is so large that inflationary fluctuations do not restore the PQ symmetry.
}
As a result, when the axion starts oscillating about its potential minima, domain walls form without attached cosmic strings.
Although an asymmetry in the domain population arises from the arbitrariness of the mean field value, the domain-wall network is robust against this population bias because of the scale-invariant nature of the initial fluctuations~\cite{Gonzalez:2022mcx}.
The domain walls can therefore enter a scaling regime even with such inflationary initial conditions~\cite{Gonzalez:2022mcx,Kitajima:2023kzu}. If they are sufficiently long-lived, the usual cosmological domain-wall problem~\cite{Zeldovich:1974uw,Vilenkin:1984ib} arises unless there is a mechanism that causes them to disappear sufficiently early.

The potential bias between different vacua resolves the degeneracy and can make the domain walls collapse and disappear.
Some of the present authors~\cite{Kitajima:2023kzu} pointed out that this fragility still holds with the scale-invariant initial fluctuations.
Here, we introduce a small potential bias term,
\begin{align}
    V_{\rm b}(\phi) 
    =
    \epsilon m_\phi^2 f_\phi^2 \left[ 1 - \cos\left(\frac{\phi}{f_{\rm b}} - \theta_{\rm b}\right) \right],
\end{align}
where $\epsilon$ is a small dimensionless parameter characterizing the size of the bias, $f_{\rm b}$ is another decay constant that satisfies $f_{\rm b} \gg f_\phi$, and $\theta_{\rm b}$ is a constant phase in the range of $[0, \pi f_\phi / f_{\rm b}]$ without loss of generality.%
\footnote{
    We can also consider a quadratic bias so that only one vacuum becomes the true vacuum in the entire field space.
    This situation can be effectively realized by taking a sufficiently large $f_{\rm b}$ compared to $H_{\rm inf}$ and $f_\phi$.
}
When $f_\phi/f_\mathrm{b}$ is rational, i.e., $f_\phi/f_\mathrm{b} = p/q$ with coprime integers $p$ and $q$, the total potential $V + V_\mathrm{b}$ has a period of $2\pi F$, where $F \equiv q f_\phi = p f_\mathrm{b}$.
Since $f_\mathrm{b} \gg f_\phi$, we have $q \gg p$, and $F$ is at least as large as $f_\mathrm{b}$.
The overall fate of the domain walls can be classified into two scenarios depending on the parameters:
\paragraph{%
Single-Bias-Period Regime
}
When the initial fluctuations are smaller than the period of the bias term but much larger than the period of $V(\phi)$, $f_\phi \ll H_{\rm inf} \lesssim f_\mathrm{b}$,
they populate multiple minima of $V(\phi)$ within a single period of the bias term.
The domain walls eventually collapse and disappear, with the field settling into the lowest-energy minimum among the populated minima~\cite{Higaki:2016yqk}.
This minimum does not necessarily coincide with the global true vacuum of the full potential, and any subsequent transition to a lower-energy minimum outside the populated range is expected to be highly suppressed in the parameter region of interest.
Thus, the dynamically selected minimum effectively becomes the vacuum of our universe.
\paragraph{%
Multiple-Bias-Period Regime
}
When the initial fluctuations are larger than the period of
the bias potential, i.e., $H_{\rm inf} \gtrsim f_{\rm b}$,
they can populate minima over multiple periods of the bias potential.
In particular, when $H_{\rm inf} \gtrsim F$,
multiple energetically equivalent lowest-energy minima, related by the periodicity of the total potential, can be populated in different Hubble patches.
As a result, the bias term does not select a unique vacuum over the populated field range, and the domain-wall networks are expected to survive for a longer time~\cite{Gonzalez:2022mcx,Kitajima:2023kzu}. 
On the other hand, when $f_{\rm b} \lesssim H_{\rm inf} \lesssim F$, the populated field range contains a unique lowest-energy minimum, and then the domain-wall network eventually disappears.
However, the lowest-energy minima in different periods of the bias potential are nearly degenerate, and the complete collapse of the wall network can be substantially delayed due to a weak bias pressure on the domain walls separating the nearly degenerate minima.

In the following sections, we study the evolution and fate of domain walls in each scenario and discuss their implications for axion dark matter isocurvature perturbations.
Before proceeding, let us clarify which contribution to the axion abundance we focus on.
When the initial fluctuations populate many minima of $V(\phi)$, the conventional axion abundance produced by the misalignment mechanism is averaged over many vacua, and its long-wavelength isocurvature perturbation can be suppressed due to the periodic nature of the potential~\cite{Kofman:1985zx,Kofman:1986wm,Linde:1990yj}.
In this paper, we do not assume that this misalignment contribution dominates the axion abundance.
Instead, we focus on axions produced by the collapse of the domain-wall network, in particular the component sourced by the biased vacuum energy released during annihilation.
In fact, if the bias is sufficiently small for the domain-wall network to enter the scaling regime before its collapse, this domain-wall contribution likely dominates over the conventional misalignment contribution.
Our central point is that the released bias energy inherits the superhorizon correlations of the inflationary initial fluctuations and transfers them to the axion energy density through the local domain-wall annihilation process.

%%%%%%%%%%%%%%%%%%%%%%%%%%%%%%%%%%%%%%
\section{Fate of Domain Walls}
\label{sec:fate-of-DWs}
%%%%%%%%%%%%%%%%%%%%%%%%%%%%%%%%%%%%%%

In this section, we examine how the macroscopic dynamics of domain walls depends on the potential and the initial fluctuations.

%%%%%%%%%%%%%%%%%%%%%%%%%%%%%%%%%%%%%%
\subsection{
Single-Bias-Period Regime
}
\label{subsec:single-bias-period}
%%%%%%%%%%%%%%%%%%%%%%%%%%%%%%%%%%%%%%

\paragraph{Bias-driven collapse.}
In this regime, the initial fluctuations populate multiple minima of $V(\phi)$ within a single period of the bias term.
The bias lifts the degeneracy among these minima and induces a pressure force on the domain walls, driving them to enlarge the regions of lower-energy vacua.
For a domain wall separating two neighboring local minima, the pressure force per unit area is set by the corresponding vacuum-energy difference, $\Delta V_i$, which depends on the pair of minima.
By contrast, the domain wall tensions are expected to be comparable among different walls if the bias term is small.
Then, walls separating vacua with larger energy differences are accelerated earlier.
Once the pressure force becomes larger than the curvature pressure due to the wall tension, which is of order $\sigma/R$, the domains of higher-energy vacua start to collapse, and the domain wall network decays.
Here, $\sigma$ is the domain wall tension, and $R$ denotes the typical curvature radius of the walls, which is of order $H^{-1}$ in the scaling regime.
This decay can proceed hierarchically: some domains disappear earlier than others, and neighboring walls may transiently approach each other or form effectively composite configurations as intermediate vacua are eliminated.

\paragraph{Selection of the endpoint vacuum.}
The classical domain-wall dynamics eventually selects a lower-energy vacuum among the populated minima.
However, the timescale for this selection depends not only on the energy differences between neighboring vacua, but also on the abundance and spatial distribution of the lower-energy domains.
The relevant question is not whether the global true vacuum of the full potential is populated, but where the lowest-energy minimum among the populated minima lies within the initial field distribution.\footnote{
A lower-energy vacuum outside the populated range may be reached only through quantum tunneling. Although bubbles of this vacuum would expand and eventually drive the universe to the global minimum, tunneling is expected to be highly suppressed in the light-axion, large-decay-constant regime of interest.
}
If this minimum lies near the center of the distribution, it is expected to be populated with an appreciable volume fraction, and the subsequent evolution is well described by the usual bias-driven collapse of the domain-wall network.
In this case, the annihilation time can be estimated by comparing the pressure force from the bias with the curvature pressure due to the wall tension.

By contrast, if the lowest-energy populated minimum lies  in the tail of the initial distribution, it may be realized only in rare regions.
Such regions may form isolated closed domains, or may not appear at all within the finite volume under consideration.
If these rare domains are subhorizon, they can collapse before they expand and percolate.
In that case, the classical domain-wall dynamics first selects the lowest-energy minimum that is actually realized in sufficiently extended regions, rather than the absolute lowest minimum within the formal populated range.
Therefore, when the lowest-energy populated minimum is a rare tail configuration, the lifetime of the domain-wall network can deviate significantly from the standard estimate.
The network may survive until a rare low-energy domain grows large enough to take over the volume, and during this period the energy density of the domain walls can become large enough to cause a domain-wall problem. At the same time, the resulting gravitational-wave signal can be enhanced, with its peak frequency shifted to lower values.
This possibility is specific to inflationary initial fluctuations, for which the field distribution can be coherent over superhorizon scales.
It is much less relevant for white-noise initial conditions, where the populated vacua are sampled locally and the lowest-energy populated vacuum is not expected to be confined only to rare, isolated regions\footnote{
In the infinite-volume limit, even such a rare vacuum occupies an infinite total volume and can be realized in regions extending over superhorizon scales, although its volume fraction is very small.
}

\paragraph{Persistence of isocurvature perturbations.}
Even after the domain wall network annihilates, the long-wavelength density fluctuations associated with the biased vacuum energy are not necessarily erased.
During the decay process, the energy stored in the domain wall network and in the biased vacuum regions is transferred to the kinetic and gradient energies of the domain walls and is subsequently converted mainly into axions, with a subdominant fraction emitted as gravitational waves.
This conversion occurs locally on the scale of the Hubble radius at the time of annihilation.
Indeed, even for inflationary initial fluctuations, the domain-wall network is expected to have an inter-wall separation of order the Hubble radius or smaller once it enters the scaling regime.\footnote{The mean separation between walls is larger than in the case of thermal initial fluctuations only by a factor of order unity~\cite{Kitajima:2023kzu}.}
Therefore, the biased vacuum energy is converted into axion energy within horizon-sized regions, rather than being transported coherently over superhorizon distances.\footnote{However, if the endpoint vacuum lies in the tail of the distribution, so that the domain-wall network takes a long time to decay, the bias energy may be transported to scales much larger than the Hubble horizon at the time when the collapse begins.}
As we will see later, this local conversion gives rise to a characteristic peak in the dimensionless power spectrum of the axion energy-density fluctuations at $k \sim a_{\rm ann} H_{\rm ann}$, where the subscript ``ann" denotes evaluation at domain wall annihilation.

To estimate the large-scale fluctuations in the bias energy released at domain wall annihilation, we consider a long-wavelength mode with $k \ll a_{\rm ann}H_{\rm ann}$. 
The resulting axion density perturbation is determined by how 
this mode modulates the coarse-grained vacuum fractions within each Hubble-sized region at the annihilation epoch, and thereby the mean bias energy density released locally during the collapse.
To estimate these fractions, we use the distribution of the initial field smoothed on the Hubble scale at annihilation. Although subhorizon isolated domains may disappear during the scaling evolution, scale-invariant fluctuations realize the same vacua in domains coherent on larger scales. Thus, in a sufficiently large region, their mean occupation fractions retain the statistical distribution inherited from the initial field fluctuations.

At domain wall formation, the axion rolls down to its nearest local minimum. When the field fluctuation spans many periods of the potential,   multiple vacua and domain walls may coexist within one Hubble volume.
After formation, the domain wall network enters the scaling regime, 
during which sufficiently small subhorizon domains are eliminated and the field configuration is progressively smoothed on short scales.
It was shown in Refs.~\cite{Gonzalez:2022mcx,Kitajima:2023kzu}
that the large-scale domain structure inherited from inflationary
fluctuations survives the subsequent scaling evolution, as expected
from causality.
Thus, we assume that
the volume fractions of the different vacua within a region of size $k^{-1}$
remain approximately unchanged up to the onset of domain-wall annihilation.
We write the coarse-grained field as $\phi_{\rm cg}=\phi_L+\phi_S$, where $\phi_L$ is the field averaged over the long wavelength scale $k$ and $\phi_S$ represents the remaining fluctuations on smaller scales on top of $\phi_L$.
Here, we smooth the field on the Hubble scale at annihilation, taking $a_{\rm ann}H_{\rm ann}$ as the upper-wavenumber cutoff for the fluctuations included in $\phi_S$.
This choice is because the domain-wall correlation length and curvature radius are of order $H_{\rm ann}^{-1}$ in the scaling regime, and the bias-driven collapse proceeds causally over a comparable scale.
As the walls propagate, the released bias energy is first converted into kinetic and gradient energy of the wall configuration and is subsequently radiated into axions through wall deformation and collapse.
We therefore assume that the axion yield on scales much larger than the Hubble radius is determined primarily by the Hubble-scale coarse-grained vacuum fractions, while the detailed subhorizon wall structure affects only the local conversion process.
As we will see shortly, the dependence of the mean bias energy on $\phi_L$ determines the power spectrum of the axion density perturbations generated at domain wall annihilation.

Let $p_n(\phi_L)$ be the conditional volume fraction, or equivalently, the conditional probability, that the field occupies the $n$-th minimum
\begin{align}
    \phi_n \simeq 2\pi n f_\phi ,
\end{align}
given the long-wavelength field value $\phi_L$.
Since $\phi_S$ represents inflationary fluctuations accumulated over the scales between $k$ and $a_\mathrm{ann} H_\mathrm{ann}$, it follows the Gaussian probability distribution,
\begin{align}
 G(\phi_S)
 =
 \frac{1}{\sqrt{2\pi\sigma_S^2}}
 \exp\left[-\frac{\phi_S^2}{2\sigma_S^2}\right], 
\end{align}
where the variance $\sigma_S^2$ is related to the Hubble parameter during inflation as
\begin{align}
    \sigma_S & = \frac{H_{\rm inf}}{2\pi} \sqrt{N_S}
\end{align}
with $N_S = \ln\left(\frac{a_{\rm ann}H_{\rm ann}}{k}\right)$.
In the limit where the bias term is small, the basin of attraction of the $n$-th minimum may be approximated by
\begin{align}
 (2n-1)\pi f_\phi
 <
 \phi
 <
 (2n+1)\pi f_\phi .
\end{align}
Then the population fraction is estimated as
\begin{align}
 p_n(\phi_L)
 &\simeq
 \int_{(2n-1)\pi f_\phi-\phi_L}^{(2n+1)\pi f_\phi-\phi_L}
 d\phi_S\,
 G(\phi_S)
 \notag\\
 &=
 \frac{1}{2}
 \left[
 {\rm erf}\left(
 \frac{(2n+1)\pi f_\phi-\phi_L}{\sqrt{2}\sigma_S}
 \right)
 -
 {\rm erf}\left(
 \frac{(2n-1)\pi f_\phi-\phi_L}{\sqrt{2}\sigma_S}
 \right)
 \right].
\end{align}
This expression shows explicitly that a long-wavelength fluctuation in $\phi_L$ shifts the local distribution of populated vacua.

With this notation, the coarse-grained biased vacuum energy density available for release during domain-wall annihilation is estimated as
\begin{align}
 \bar\rho_{\rm b}(\phi_L)
 =
 \sum_n p_n(\phi_L)
 \left[
 V_{\rm b}(\phi_n)-V_{\rm b}(\phi_{\rm end})
 \right],
 \label{eq:rhob_coarse}
\end{align}
where $\phi_{\rm end}$ is the endpoint vacuum selected by the classical domain-wall dynamics.
The subtraction of $V_{\rm b}(\phi_{\rm end})$ fixes the zero of the released vacuum energy.
The quantity $\bar\rho_{\rm b}(\phi_L)$ should be understood as a coarse-grained average over many Hubble-sized regions inside the scale $k^{-1}$.

A long-wavelength fluctuation therefore modulates the released energy density as
\begin{align}
 \delta\rho_{\rm b}({\bf x})
 \simeq
 \frac{\partial \bar\rho_{\rm b}}{\partial \phi_L}
 \delta\phi_L({\bf x})
 +
 \frac{1}{2}
 \frac{\partial^2 \bar\rho_{\rm b}}{\partial \phi_L^2}
 \left[
  [\delta\phi_L({\bf x})]^2
  -
  \left\langle (\delta\phi_L)^2 \right\rangle
 \right],
\end{align}
where the derivatives are evaluated at the background value $\phi_L=\langle\phi\rangle$.
For Gaussian axion fluctuations, the cross correlation between the linear and quadratic terms vanishes.
Thus, for $k \ll a_{\rm ann}H_{\rm ann}$, the dimensionless power spectrum for the fractional perturbation $\delta_{\rm b}\equiv\delta\rho_{\rm b}/\bar\rho_{\rm b}$ is given by
\begin{align}
 \Delta_{\delta_{\rm b}}^2(k)
 \simeq
 \left(
  \frac{\partial\ln\bar\rho_{\rm b}}{\partial \phi_L}
 \right)^2
 \Delta_\phi^2(k)
 +
 \frac{1}{4}
 \left(
  \frac{1}{\bar\rho_{\rm b}}
  \frac{\partial^2\bar\rho_{\rm b}}{\partial \phi_L^2}
 \right)^2
 \Delta_{(\delta\phi_L)^2}^2(k),
\end{align}
where $\Delta_{(\delta\phi_L)^2}^2$ denotes the dimensionless power spectrum of $(\delta\phi_L)^2-\langle(\delta\phi_L)^2\rangle$.\footnote{For Gaussian fluctuations, its dimensional power spectrum is $P_{(\delta\phi_L)^2}(k)=2\int d^d q/(2\pi)^d\,P_\phi(q)P_\phi(|{\bf k}-{\bf q}|)$, where $d$ is the number of spatial dimensions.}
Therefore, as long as the linear response $\partial\bar\rho_{\rm b}/\partial\phi_L$ is not accidentally suppressed, the first term dominates and the biased vacuum energy inherits the nearly scale-invariant spectrum of the inflationary axion fluctuations on superhorizon scales.
If the linear response is suppressed, for example by symmetry at $\langle\phi\rangle=0$, the quadratic term instead gives the leading nonzero contribution.

\begin{figure}[tbp]
\centering 
\includegraphics[width=7.5cm,clip]{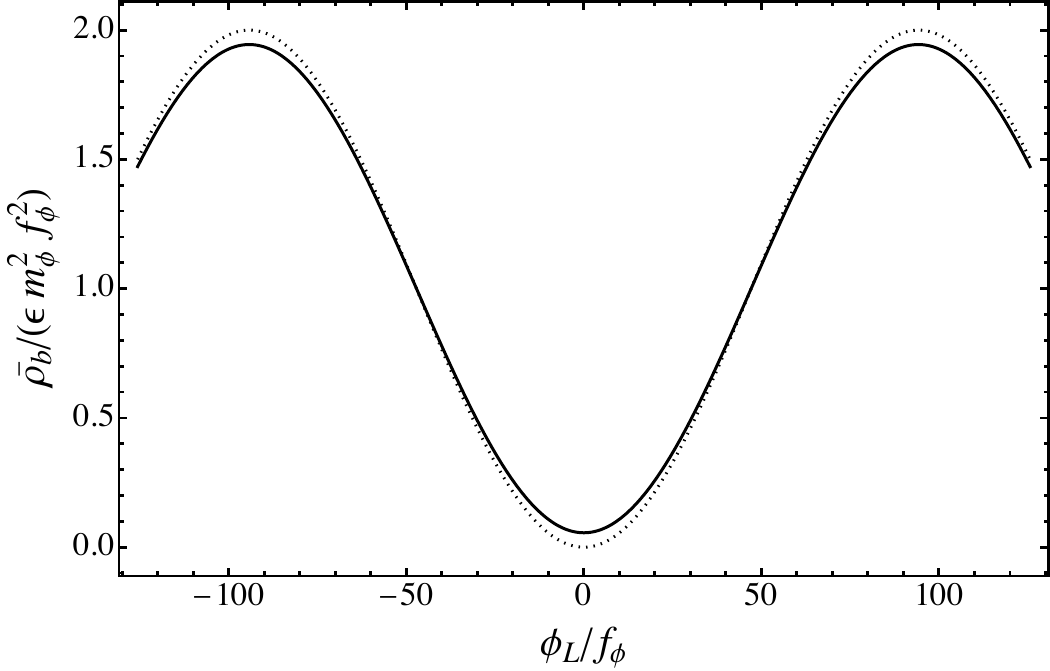}
\includegraphics[width=7.5cm,clip]{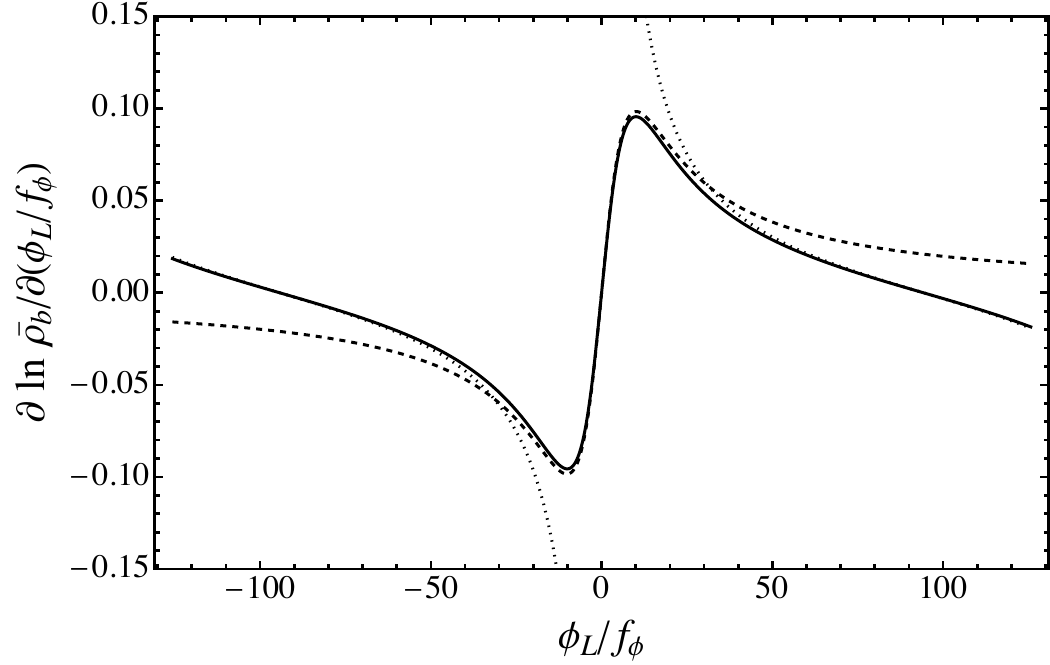}
\caption{
Coarse-grained released bias energy and its linear response for
$f_\mathrm{b}/f_\phi=30$, $\sigma_S/f_\phi=10$, $\theta_\mathrm{b}=0$, and $\phi_{\rm end}=0$.
Left: $\bar\rho_{\rm b}/(\epsilon m_\phi^2 f_\phi^2)$ as a function of
$\phi_L/f_\phi$.
Right: $\partial \ln \bar\rho_{\rm b}/\partial(\phi_L/f_\phi)$.
The solid line is the discrete-sum result for the periodic bias potential,
the dotted line is the bare periodic bias potential evaluated at $\phi=\phi_L$,
and the dashed line is the discrete-sum result for the quadratic mass-term
bias obtained in the large-$f_\mathrm{b}$ limit.
}
\label{fig: linear response}
\end{figure}

In Fig.~\ref{fig: linear response}, we plot $\bar{\rho}_{\rm b}(\phi_L)$ and 
$\partial\ln\bar\rho_{\rm b}/\partial(\phi_L/f_\phi)$ as functions of $\phi_L/f_\phi$.
For the solid line, we take $f_{\rm b}=30 f_\phi$, $\sigma_S=10 f_\phi$, 
$\theta_{\rm b}=0$, and $\phi_{\rm end}=0$.
For comparison, we also show the bare bias potential as a dotted line.
The solid line does not vanish at $\phi_L=0$, because the coarse-grained
energy receives contributions from the minima around the origin.
In the right panel, the linear response function shown by the solid line is
generically nonzero, except at the origin and $\phi_L/f_\phi = \pm 30\pi$.
We also show the large-$f_\mathrm{b}$ limit as a dashed line, in which the bias potential is effectively
quadratic. In this case, the linear response coefficient vanishes only at the
origin.

In this figure, we have fixed $\phi_{\rm end}=0$. As $|\phi_L|$ increases, 
the initial volume fraction of the vacuum $\phi_{\rm end}=0$ becomes
small. In particular, for the adopted parameters, $\phi_{\rm end}=0$ lies
outside the $2\sigma_S$ range of the Gaussian distribution of the
short-wavelength fluctuations when $|\phi_L|\gtrsim 20 f_\phi$.
Therefore, some care is needed when interpreting the figure at large
$|\phi_L|$.

Since domain-wall annihilation is a causal process occurring on horizon or subhorizon scales, it cannot smooth out fluctuations whose wavelengths are superhorizon at the time of annihilation.
Consequently, the corresponding isocurvature perturbations on superhorizon scales can survive the domain-wall dynamics.
Equivalently, the annihilation process may be regarded as inducing a local transfer function that maps fluctuations in the biased vacuum energy to those in the axion energy density.
For modes with $k \ll a_{\rm ann}H_{\rm ann}$, this transfer function is expected to approach a nearly scale-independent response, leaving the large-scale spectral tilt essentially unchanged.
Thus, the axion energy-density spectrum retains a nearly scale-invariant component inherited from the superhorizon fluctuations of the bias energy:
\begin{align}
\Delta_S^2(k) \simeq r_{\rm b}^2 \Delta_{\delta_{\rm b}}^2(k)
\end{align}
where $\Delta_S^2(k)$ denotes the dark matter isocurvature power spectrum, $r_{\rm b}$ denotes the fraction of the bias contribution to the final axion density, and here we assume that the axion dark matter produced by domain-wall annihilation accounts for the total dark matter density.
As we will see later, $r_{\rm b}$ is order unity.

%%%%%%%%%%%%%%%%%%%%%%%%%%%%%%%%%%%%%%
\subsection{
Multiple-Bias-Period Regime
}
\label{subsec: multiple vac}
%%%%%%%%%%%%%%%%%%%%%%%%%%%%%%%%%%%%%%

First, we discuss the case of $H_{\rm inf} \gtrsim F$.
In this case, the inflationary fluctuations are larger than the period of the total potential, and multiple true vacua are populated after the formation of domain walls.
When the potential bias overcomes the domain-wall tension, domain walls are driven to expand the regions with lower potential energy.
As a result, most of the universe is eventually filled with the degenerate true vacua.
At the interfaces between adjacent true-vacuum regions, the field value must pass through multiple local minima and maxima of the potential.
Such a field configuration can be regarded as a composite wall made of multiple elementary domain walls, each of which separates adjacent local minima of the potential.
The subsequent evolution is therefore essentially the same as that of a domain-wall network formed by the bias potential itself.
In particular,  such walls with initial inflationary fluctuations are expected to be highly robust against a population bias~\cite{Gonzalez:2022mcx,Kitajima:2023kzu}.
Since the true vacua remain degenerate, these composite walls do not feel any pressure from the bias potential.
Therefore, once such walls are formed, the resulting domain-wall network becomes long-lived and may lead to the domain-wall problem unless the remaining degeneracy is lifted by an additional potential bias.

Let us briefly discuss the profile of this composite domain wall.
For adjacent true vacua, one has
\begin{align}
    \Delta \phi_\mathrm{true}
    =
    2\pi F
    =
    2\pi q f_\phi
    .
\end{align}
A field configuration connecting them must pass through the intervening local minima and maxima of $V( \phi ) + V_\mathrm{b} ( \phi )$.
This is a composite, or sandwich, wall made of $q$ elementary walls, which is in close analogy with the wall-sandwich picture discussed in Ref.~\cite{Gabadadze:2000vw}.
For an elementary wall associated with $V( \phi )$, its width and tension of each wall, $\delta_0$ and $\sigma_0$, respectively, are estimated as
\begin{align}
    \delta_0 \sim m_\phi^{-1}
    \qquad
    \sigma_0 \simeq 8m_\phi f_\phi^2
    .
\end{align}

The width of the composite wall $\ell_\mathrm{comp}$ can be estimated as follows.
Each elementary core has a width of order $m_\phi^{-1}$, while the separations between neighboring cores are controlled by the competition between the exponentially weak kink-kink repulsion and the small energy cost of the intermediate plateau induced by the bias term.
For the benchmark case $p=1$ and $\theta_\mathrm{b} = 0$, for which $q = f_\mathrm{b} / f_\phi$, one finds parametrically
\begin{align}
    \ell_\mathrm{comp}
    \sim
    \frac{q}{m_\phi} \log \frac{1}{\epsilon}
    ,
\end{align}
where the logarithmic factor comes from the balance between the Yukawa-like repulsion force and the adjacent vacuum energy difference.

For small $\epsilon$, the elementary walls are separated by sufficiently large distances, and the leading estimate for the composite-wall tension is simply additive,
\begin{align}
    \sigma_\mathrm{comp}
    \simeq
    q\sigma_0
    \simeq
    8 q m_\phi f_\phi^2
    .
\end{align}
The order-unity corrections neglected in this estimate depend on the detailed multi-wall profile and on the relative phase $\theta_b$.

Next, we discuss the case of $f_{\rm b} \lesssim H_{\rm inf} \lesssim F$.
In this case, the populated field range does not span a full period of the total potential and therefore contains a unique lowest-energy minimum.
Thus, we expect the domain wall network to eventually disappear.
However, due to the small energy differences between local energy minima in different periods of the bias potential, the complete disappearance of the network can be substantially delayed from the onset of the network collapse.

To see this delayed collapse, let us estimate the energy differences between the local minima of the total potential.
Since the bias potential is subdominant in the total potential, the local minima are approximately determined by the minima of $V$, $\phi_n = 2\pi n f_\phi$.
Then, the energy difference between two adjacent local minima is estimated as 
\begin{align}
    \Delta V_{{\rm adj},n}
    &\simeq
    |V_{\rm b}(\phi_{n+1}) - V_{\rm b}(\phi_n)|
    =
    \epsilon m_\phi^2 f_\phi^2 
    \left|
        \cos \left( \frac{2\pi n f_\phi}{f_{\rm b}} - \theta_{\rm b} \right)
        -
        \cos \left( \frac{2\pi (n+1)f_\phi}{f_{\rm b}}  - \theta_{\rm b} \right)
    \right|
    \nonumber \\
    &\simeq 
    \epsilon m_\phi^2 f_\phi^2 
    \frac{2\pi f_\phi}{f_{\rm b}} 
    \left| \sin \left( \frac{2\pi n f_\phi}{f_{\rm b}}  - \theta_{\rm b}\right) \right|
    ,
\end{align}
where we used $f_\phi/f_{\rm b} \ll 1$.
The magnitude of this energy difference depends strongly on $n$ through the sine factor.
Around the inflection points of $V_{\rm b}$, it can be as large as
\begin{align}
    \Delta V_{{\rm adj},n}
    \sim
    2\pi\epsilon m_\phi^2\frac{f_\phi^3}{f_{\rm b}}.
\end{align}
On the other hand, near extrema of $V_{\rm b}$, the sine factor is suppressed by a factor of $\mathcal{O}(f_\phi/f_{\rm b})$, and the energy difference becomes as smalll as
\begin{align}
    \Delta V_{{\rm adj},n}
    \sim
    2\pi\epsilon m_\phi^2\frac{f_\phi^4}{f_{\rm b}^2}.
\end{align}

As discussed above, the elementary domain walls separating adjacent minima have approximately the same tension $\sigma_0$.
Thus, domain walls associated with larger $\Delta V_{{\rm adj}, n}$ are pushed by larger pressure forces and start to move earlier.
In particular, domain walls  around the inflection points of $V_{\rm b}$ tend to disappear earlier, whereas those around the extrema of $V_{\rm b}$ can survive until later times.

At the end of this stage, the domain walls around the minima of $V_{\rm b}$ are finally pushed away, and the elementary walls combine into composite walls, as in the previous case.
These composite walls separate the lowest-energy minima in neighboring periods of the bias potential.
In contrast to the previous case, these minima are not exactly degenerate.
Therefore, the composite walls experience a bias pressure and eventually collapse.
The energy difference relevant to the collapse of the composite walls is typically $\Delta V \sim 2\pi \epsilon m_\phi^2 f_\phi^4/f_{\rm b}^2$.
Moreover, the tension of a composite wall is much larger than that of an elementary wall, $\sigma_{\rm comp}\gg\sigma_0$.
Thus, although the domain-wall network eventually disappears, the network of composite walls can remain for a substantial period before its complete collapse.

%%%%%%%%%%%%%%%%%%%%%%%%%%%%%%%%%%%%%%
\section{Numerical Lattice Simulations}
\label{sec: lattice}
%%%%%%%%%%%%%%%%%%%%%%%%%%%%%%%%%%%%%%

We performed two-dimensional lattice simulations to verify the network evolution discussed in the previous section.
In the numerical simulations, we assume a radiation-dominated universe
and choose the initial time $\tau_i = 1$
such that $H_i=m_\phi$. Here $\tau$ is the dimensionless conformal time, and
the scale factor is normalized to unity at the initial time, $a_i=1$. 
The initial fluctuations are taken to have a scale-invariant power spectrum  over the finite range
$k_{\rm IR}<k<k_{\rm UV}$,
\begin{align}
    \mathcal{P}(k) = \Delta_\phi^2 \, \Theta(k-k_{\rm IR})\,\Theta(k_{\rm UV}-k) ,
    \label{eq:initial_fluc}
\end{align}
where $\Delta_\phi$ is an amplitude independent of $k$, and $\Theta(x)$ is the Heaviside step function.
We generate the initial scalar-field configuration as
\begin{align}
    \phi(\bf{x}) = \langle \phi \rangle + \delta \phi(\bf{x})
    ,
\end{align}
where $\langle \phi \rangle$ is a homogeneous component, and $\delta \phi$ is a Gaussian random field by sampling the Fourier coefficients from zero-mean Gaussian distributions. In two spatial dimensions, Eq.~(\ref{eq:initial_fluc}) corresponds to
$P_\phi(k)=2\pi\Delta_\phi^2/k^2$ for
$k_{\rm IR}<k<k_{\rm UV}$, with the modes outside this range set to zero.
We choose $k_{\rm IR}=2\pi/L$ and
$k_{\rm UV}=2 \pi a_i m_\phi=2\pi a_iH_i$, where $L$ is the comoving box
size.
With the power spectrum of Eq.~(\ref{eq:initial_fluc}), the variance of the field value at each grid point is given by $\Delta_\phi^2 \ln(k_\mathrm{UV}/k_\mathrm{IR}) = \Delta_\phi^2 \ln ( m_\phi L )$.
To ensure that the amplitude of fluctuations is independent of the box size $L$, we use $\Delta_\phi^2 \ln ( m_\phi L)$ as an input parameter of lattice simulations.

In the following, we focus on the single-bias-period regime.
For simplicity, we approximate the bias potential by a quadratic form as
\begin{align}
    V_{\rm b}(\phi) = \frac{1}{2}\, \tilde{\epsilon} \,m_\phi^2 \phi^2, 
\end{align}
where we set $\theta_b =0$ and $\tilde{\epsilon} \equiv \epsilon f_\phi^2/f_{\rm b}^2$.

\begin{figure}[tbp]
\centering 
\includegraphics[width=4cm,clip]{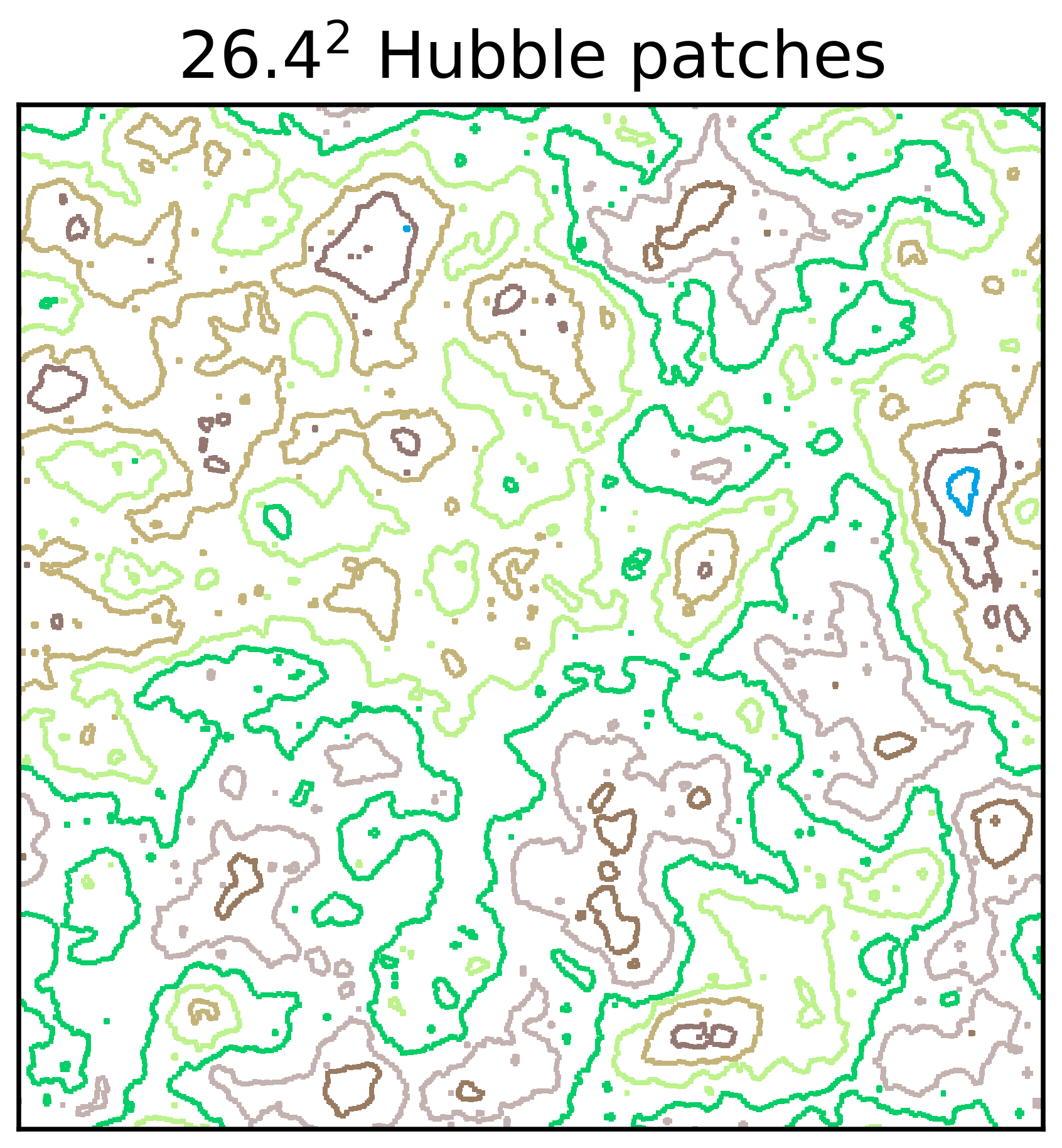}
\includegraphics[width=4cm,clip]{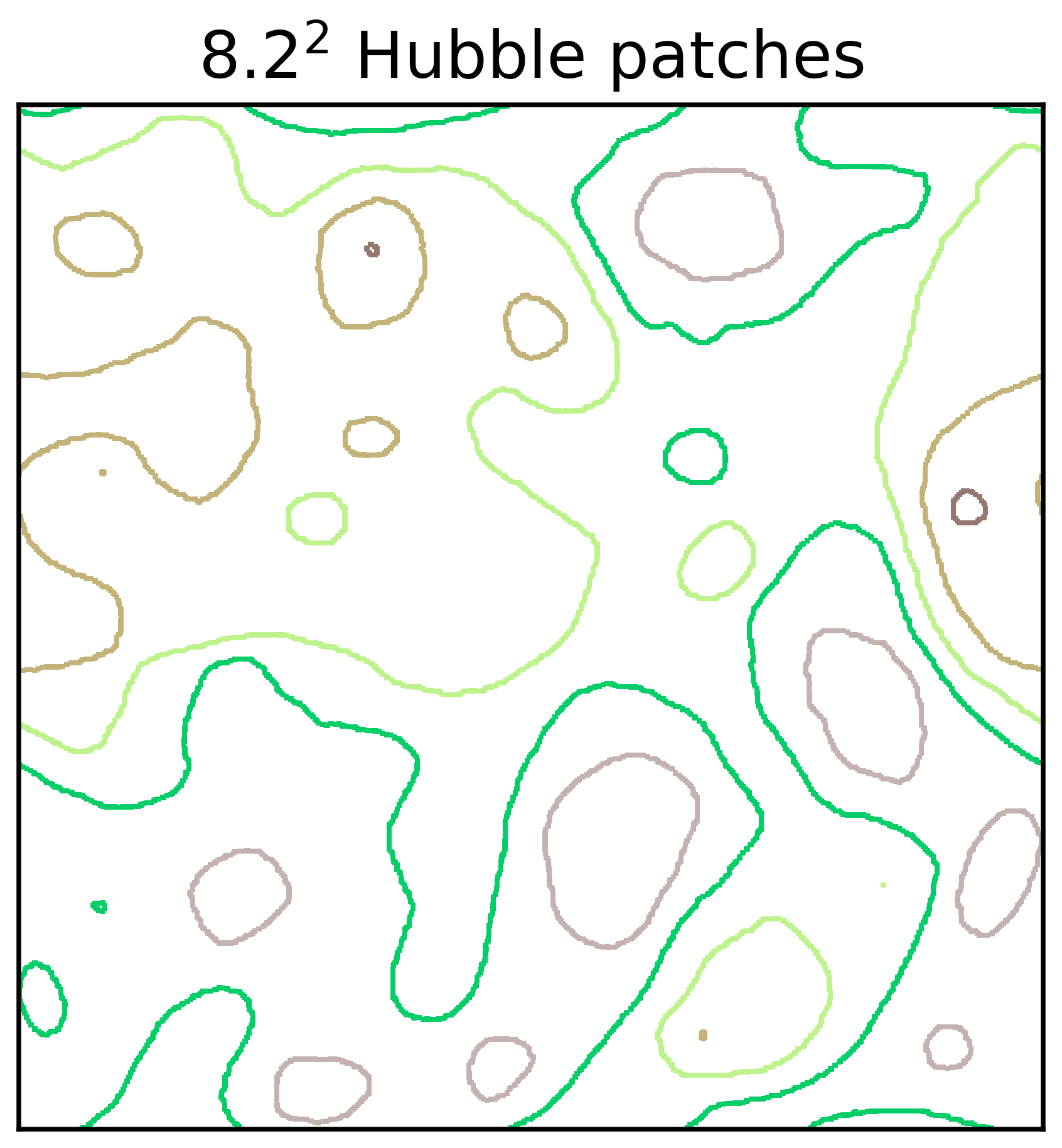}
\includegraphics[width=4cm,clip]{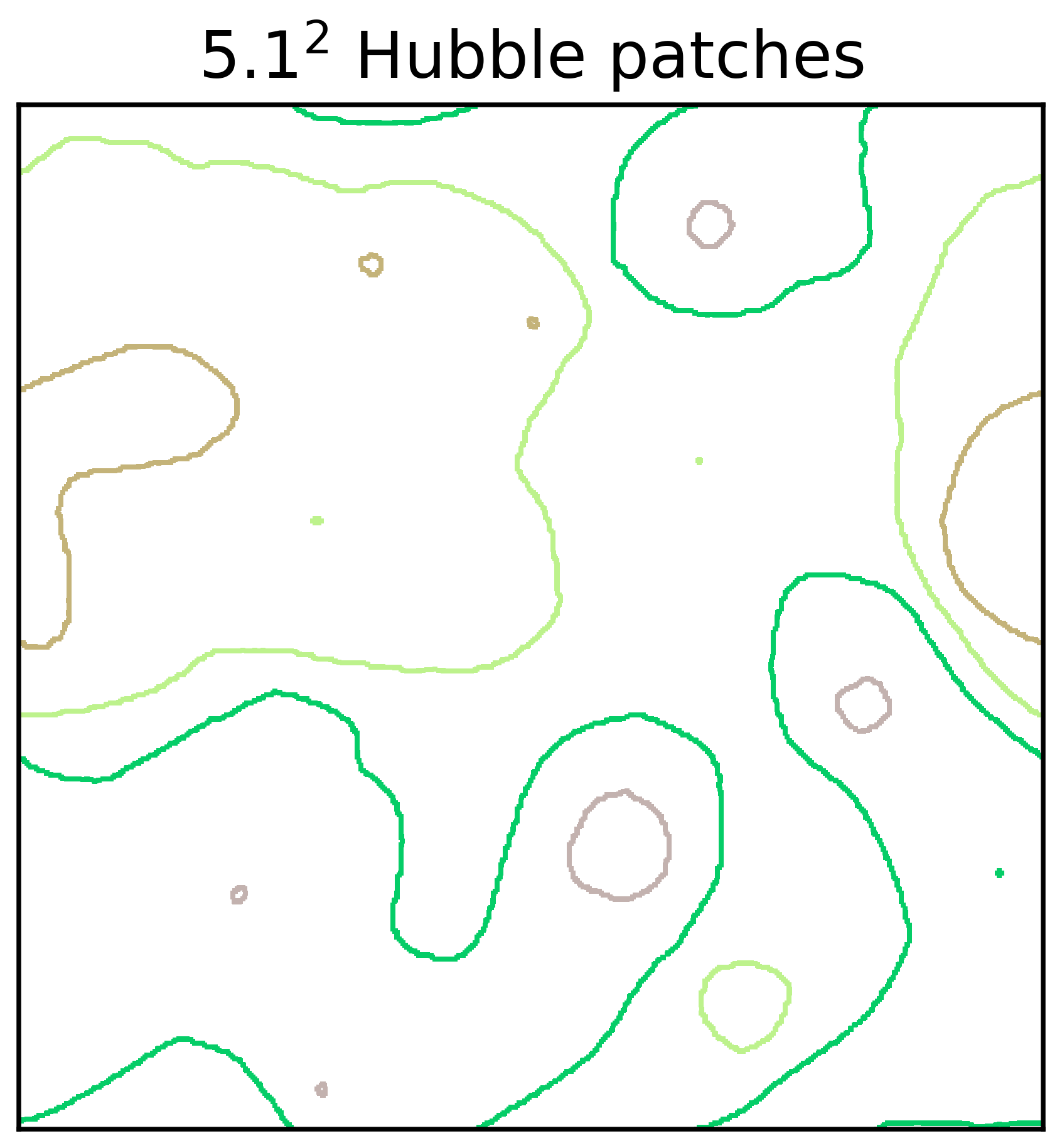}
\includegraphics[width=4cm,clip]{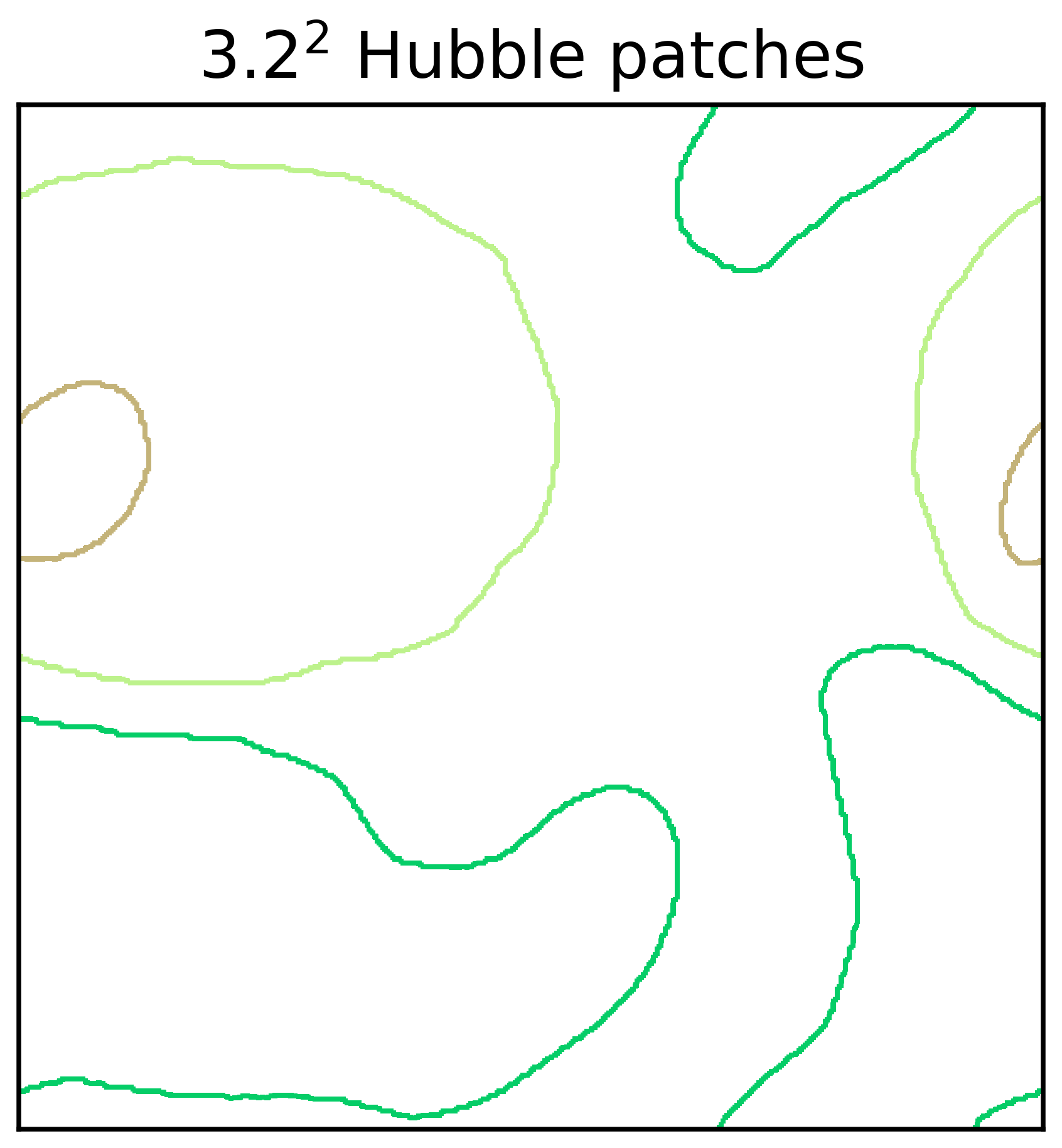}
 \caption{
 Snapshots of the domain-wall network evolution without bias in the two-dimensional lattice simulation with $16384^2$ grid points, corresponding to $132^2$ Hubble patches initially. Different colors distinguish different types of domain walls. We set 
 $\langle \phi \rangle = 0$ and
 $\Delta_\phi = 4.5\pi f_\phi/\sqrt{\ln(m_\phi L)}$. 
 The panels are arranged chronologically from left to right, and the simulation box contains $26.4^2$, $8.2^2$, $5.1^2$, and $3.2^2$ Hubble patches at the respective times.
}
\label{fig:snapshots}
\end{figure}

First, we performed a simulation without the potential bias, i.e., $\tilde{\epsilon}=0$, on a two-dimensional lattice with $16384^2$ grid points, initially covering $132^2$ Hubble patches, to confirm the scaling behavior of the domain walls for scale-invariant initial fluctuations.
We show snapshots of the domain wall network evolution in Fig.~\ref{fig:snapshots}.
Since we generate large initial fluctuations, multiple vacua are populated, 
forming several types of domain walls, which are distinguished by the color in the figure.
We can see that some domains extend beyond a single Hubble patch.
For example, in the final panel, the simulation box contains $7.8^2$ Hubble patches, while individual domains extend across several of them.
This reflects the scale-invariant nature of the initial fluctuations: correlations on superhorizon scales survive domain-wall formation and its subsequent evolution, making the network robust against the initial population bias~\cite{Gonzalez:2022mcx,Kitajima:2023kzu}.

\begin{figure}[tbp]
\centering 
\includegraphics[width=4cm,clip]{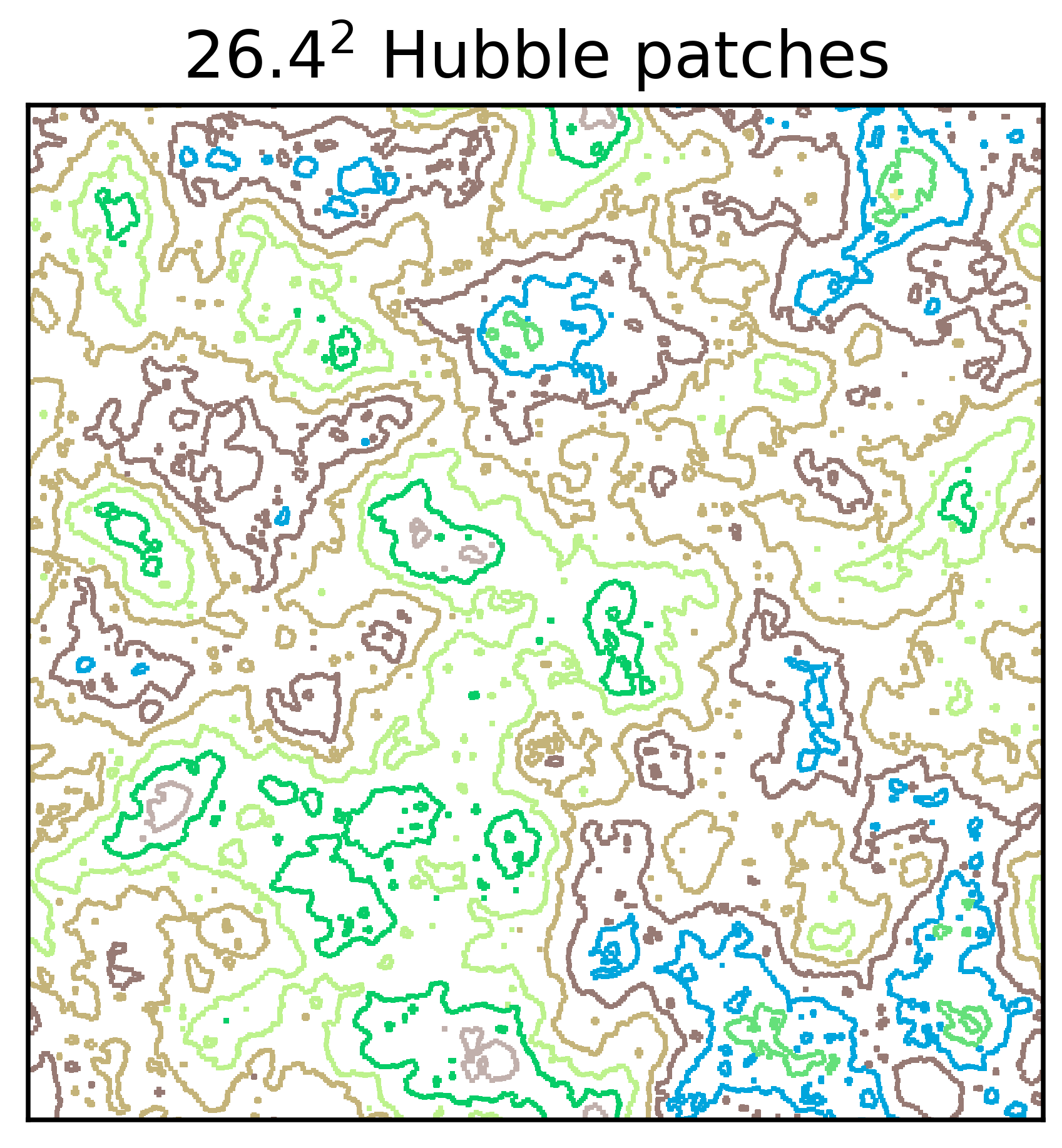}
\includegraphics[width=4cm,clip]{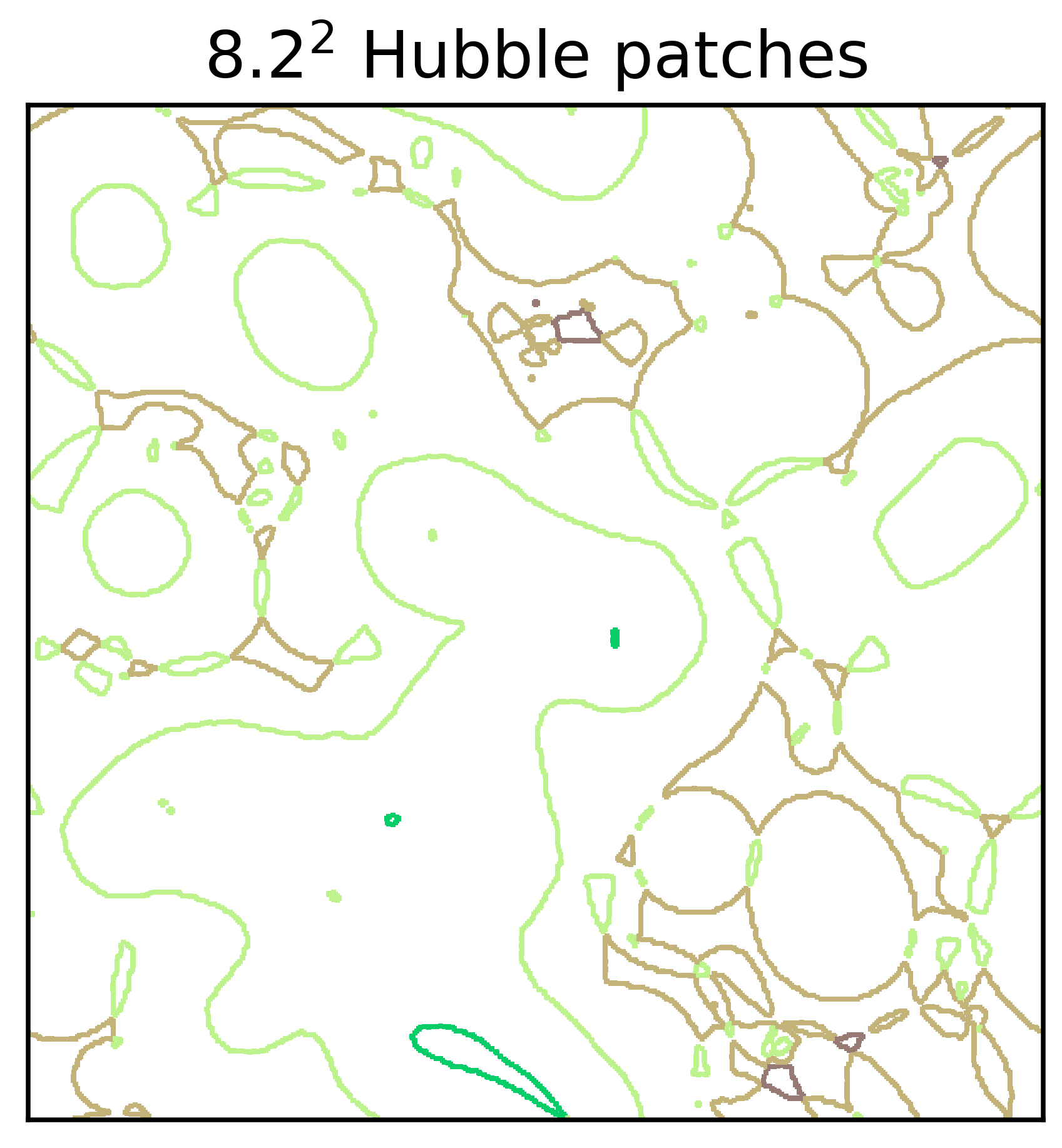}
\includegraphics[width=4cm,clip]{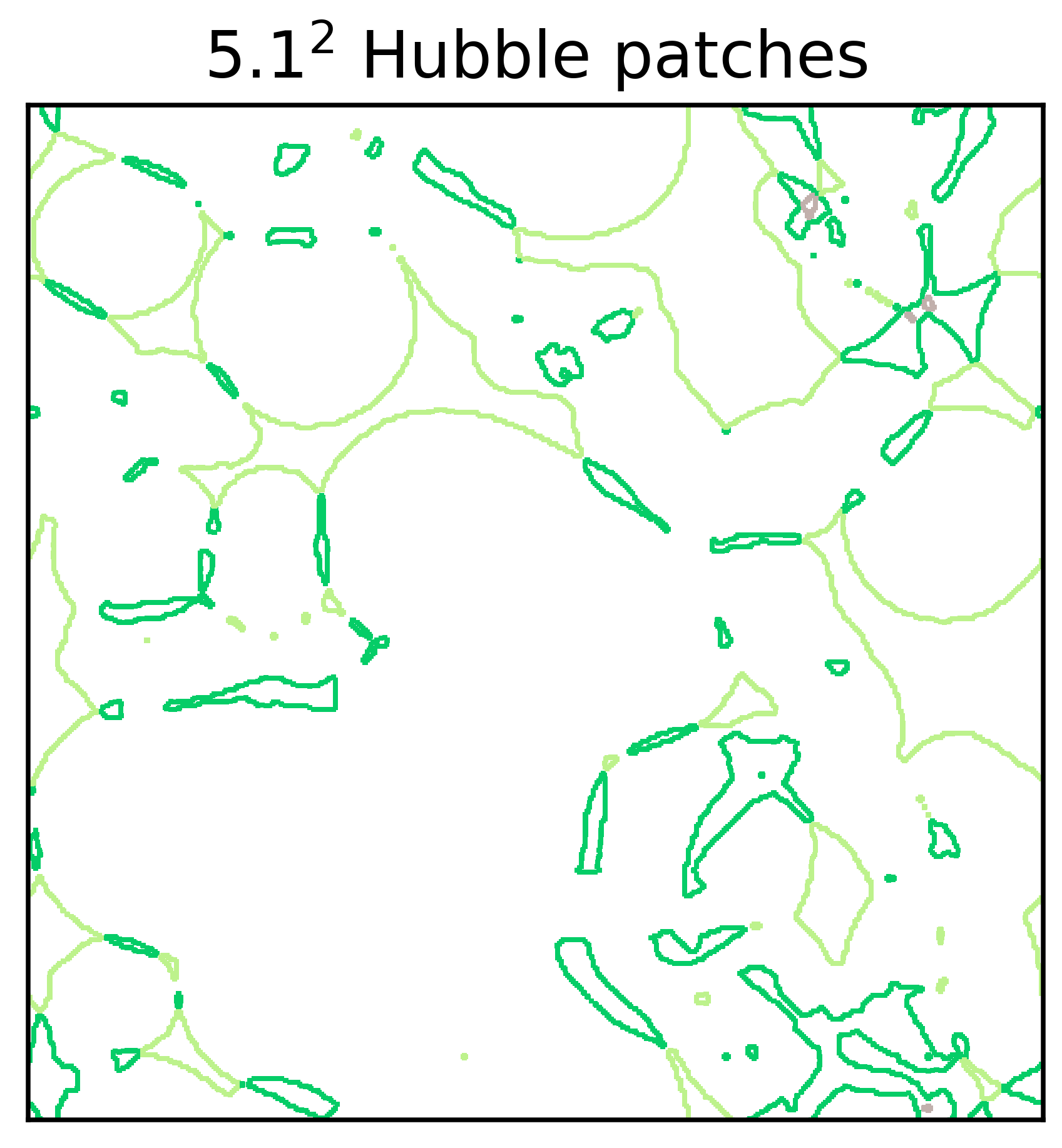}
\includegraphics[width=4cm,clip]{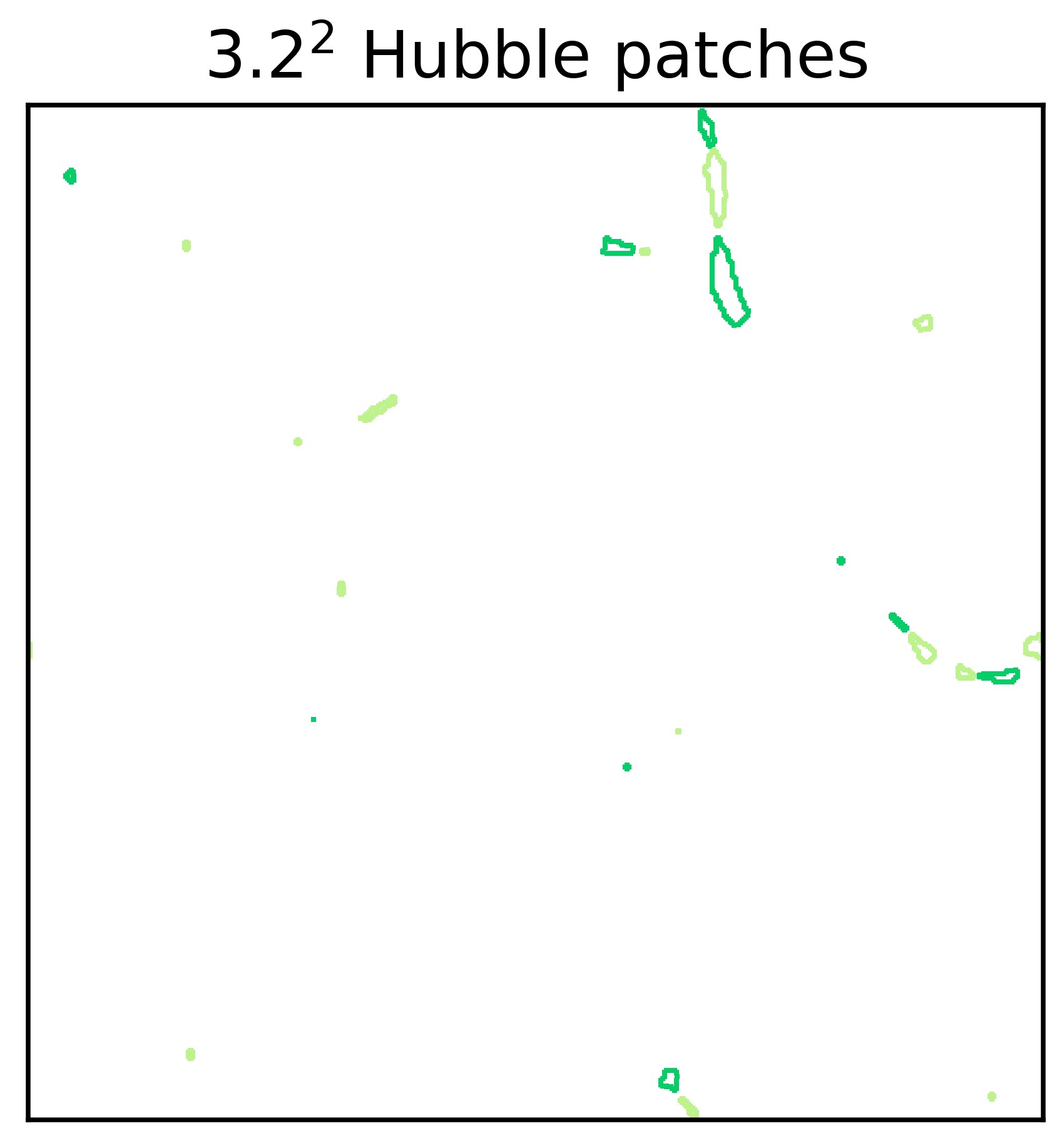}
\caption{
Same as Fig.~\ref{fig:snapshots} but with bias $\tilde{\epsilon} = 0.0056$ and $\langle \phi \rangle = 3.5 \pi f_\phi$.
At the final time slice, the space is dominated by the global minimum $\phi = 0$.
The snapshots of the field configurations and energy density distributions are shown in the Fig.~\ref{fig:rho_theta_snapshots}.
}
\label{fig:snapshots_bias}
\end{figure}

Next, we performed the simulation with the bias of $\tilde{\epsilon} = 0.0056$, using the same lattice setup.
Figure~\ref{fig:snapshots_bias} shows the corresponding evolution in the presence of a potential bias.
By the final panel, the domain walls have mostly collapsed, and the domain associated with the lowest-energy vacuum dominates the simulation volume.
Energy conservation and causality suggest that the large-scale spatial distribution of the axion energy produced during the collapse remains correlated with the vacuum-domain distribution at the time of wall formation, which in turn inherits the large-scale correlations of the initial fluctuations.

\begin{figure}[tbp]
    \centering
    \includegraphics[width=0.96\textwidth]{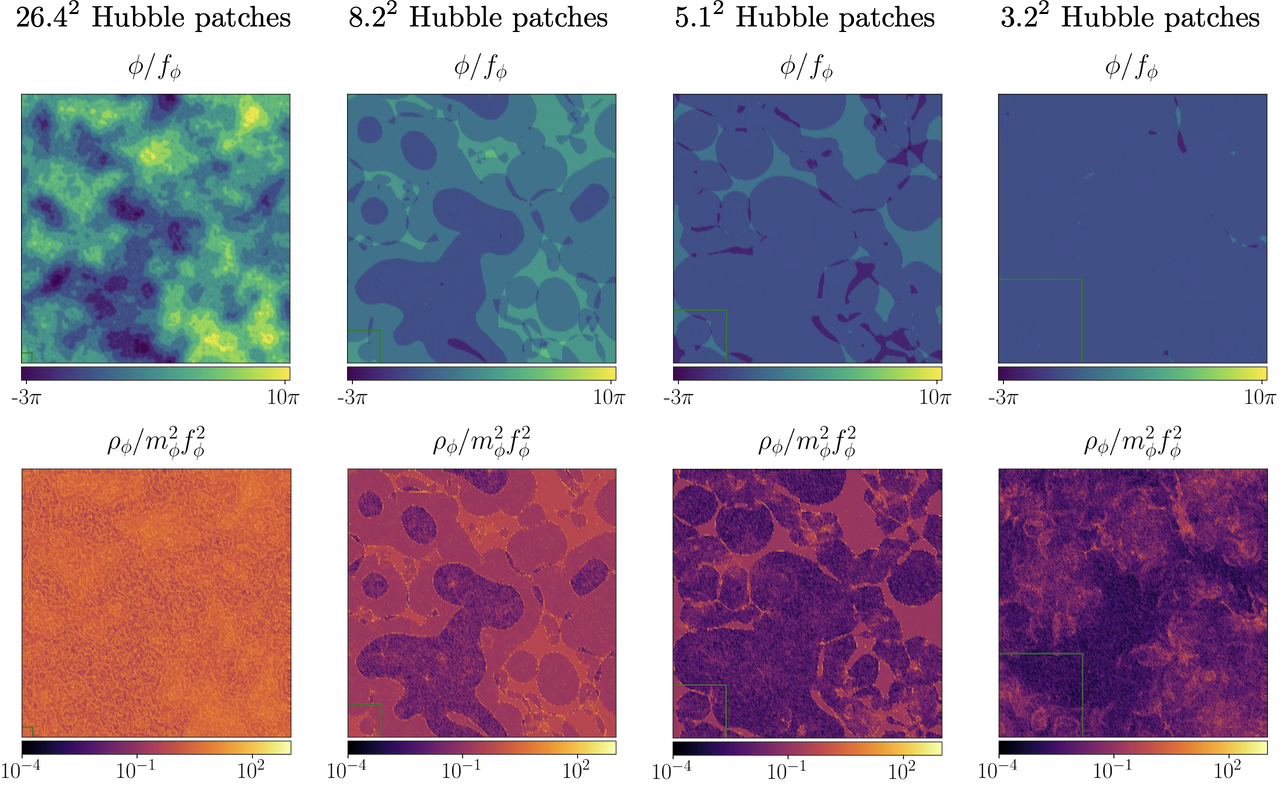}
    \caption{
    Snapshots of field configurations (\textbf{upper}) and energy density distributions (\textbf{lower}) corresponding to Fig.~\ref{fig:snapshots_bias}.
    The square framed by the green line at the bottom-left corner in each plot displays the Hubble patch at each time slice.
    Focusing on the locations of high-energy regions on the final time slice shown in the \textbf{lower right} panel, one can observe a significant overlaps with the regions occupied by higher-energy vacua in the early-time field configuration shown in the \textbf{upper left} panel.
    }
    \label{fig:rho_theta_snapshots}
\end{figure}

This behavior can also be seen more directly in the snapshots of the axion energy-density evolution shown in Fig.~\ref{fig:rho_theta_snapshots}.
The lower panels show the axion energy density, while the upper panels show the corresponding field configurations.
In the final energy-density snapshot, the high-energy regions exhibit a spatial distribution extending over scales larger than the Hubble size, indicated by the green square.
Their locations overlap significantly with the regions occupied by higher-energy vacua in the early-time field configuration.
This provides direct visual evidence that the large-scale correlations in the initial vacuum distribution are transferred to the final axion energy density.
If we look more closely, we can also see remnant structures in the axion energy density associated with the domain-wall collapse, extending over scales of order the Hubble radius.

\begin{figure}[tbp]
\centering 
\includegraphics[width=0.65\textwidth,clip]{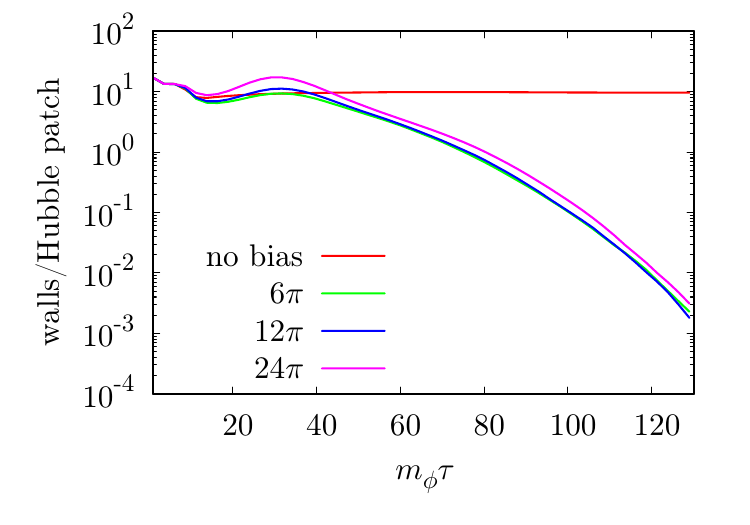}
\caption{
Evolution of the normalized domain-wall density, $\mathcal{A}/(VH)$, where $\mathcal{A}$ denotes the total physical wall length and $V$ the physical area of the simulation box.
We set $\tilde\epsilon = 0$ (red), $0.001$ (others) and $\langle \phi \rangle = 0$ (red), $6 \pi f_\phi$ (green), $12 \pi f_\phi$ (blue), $24 \pi f_\phi$ (magenta) for the initial distribution. 
The number of walls per Hubble patch is larger than in the case of small initial fluctuations.
}
\label{fig:wall_density_evolution}
\end{figure}

In Fig.~\ref{fig:wall_density_evolution}, we show the time evolution of the wall density with and without the potential bias.
Here and in the following lattice simulations, we performed simulations on a two-dimensional lattice with $524288^2$ grid points and $L = 2048/m_\phi$.
Without the potential bias, the wall density per one Hubble patch approaches a constant, which implies the scaling regime.
Unlike the case of small initial fluctuations, large fluctuations on superhorizon scales produce field excursions extending across multiple Hubble patches and spanning many vacua. The domain walls associated with these successive vacuum crossings are not efficiently removed by pairwise annihilation and therefore remain more abundant, resulting in a larger wall density than in a case of the $Z_2$ domain walls.
Even in the presence of the potential bias, the wall density also temporarily exhibits scaling behavior before rapidly decreasing as the network collapses. While the time evolution is similar regardless of the value of $\langle\phi\rangle$, the wall density itself is larger for larger $\langle\phi\rangle$.
For larger values of $\langle\phi\rangle$, the true-vacuum domains occupy a smaller fraction of the initial volume.
The walls bounding these domains must therefore propagate over greater distances before the true vacuum fills the simulation volume.
Consequently, the collapse of the network takes longer than in cases where the true vacuum initially occupies a larger volume fraction, leaving a higher wall density at a given time.

\begin{figure}[tbp]
\centering 
\includegraphics[width=0.49\textwidth,clip]{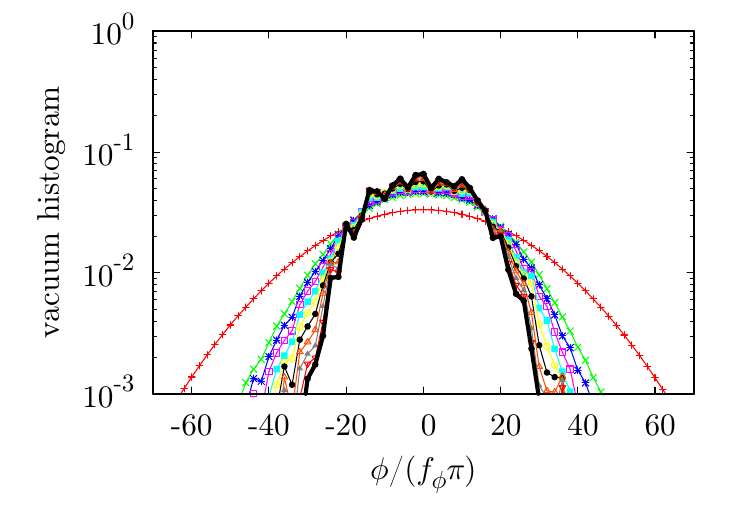}
\includegraphics[width=0.49\textwidth,clip]{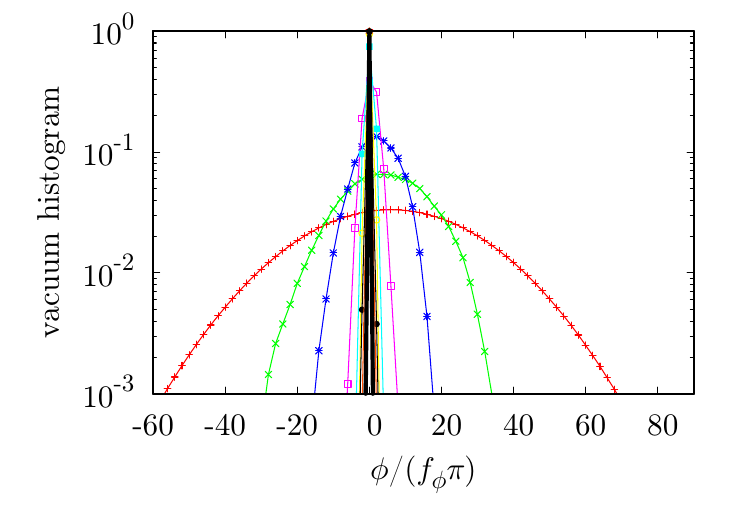}
\includegraphics[width=0.49\textwidth,clip]{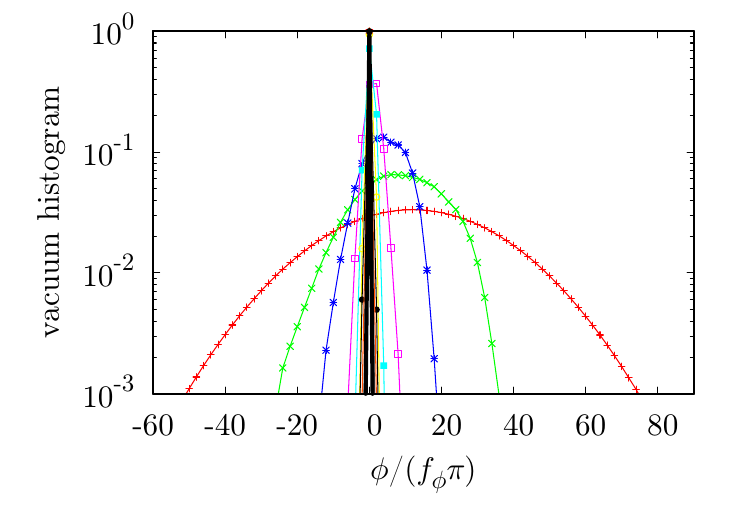}
\includegraphics[width=0.49\textwidth,clip]{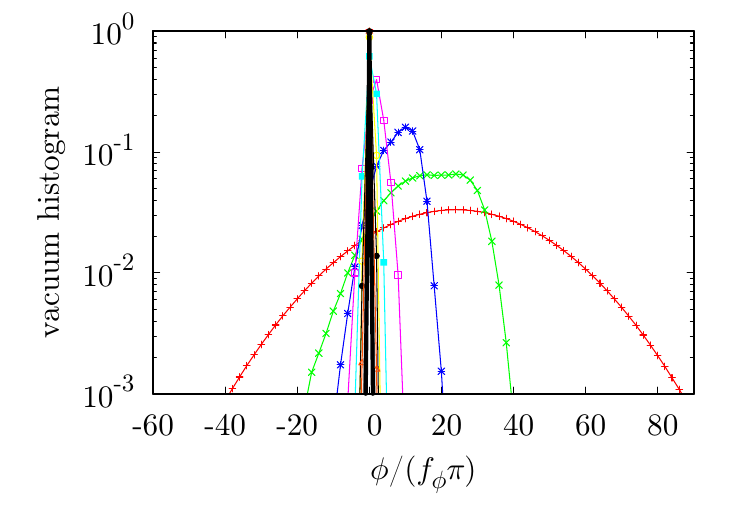}
\caption{Evolution of the vacuum histogram for $\tilde\epsilon = 0$ (top-left) and $0.001$ (others). The red and thick black points show the initial ($\tau = 1$) and final ($\tau = 129$) distribution, respectively, and the time interval between each plot is $12.8$. We set $\langle \phi \rangle = 0$ (top-left), $6 \pi f_\phi$ (top-right), $12 \pi f_\phi$ (bottom-left), $24 \pi f_\phi$ (bottom-right) for the initial distribution.
}
\label{fig:vacuum_histrogram}
\end{figure}

In Fig.~\ref{fig:vacuum_histrogram}, we show the time evolution of the field-value histogram.
Without the potential bias (top-left), while the histogram becomes narrower as time passes, the field distribution remains spread even at the final time of the simulation.
This evolution of the histogram is understood as an artifact due to the finite volume of the simulation box.
If the field fluctuations are exactly scale-invariant, even vacua corresponding to the tails of the field distribution can be populated over arbitrarily large spatial scales.
However, in a simulation box with a finite volume, such rare field values are unlikely to extend over large regions.
Consequently, they appear only as small isolated domains, which are subsequently eliminated by the domain-wall dynamics.
On the other hand, with the potential bias, the histogram rapidly evolves toward the global minimum, $\phi = 0$.
This trend is independent of the value of $\langle \phi \rangle$.

\begin{figure}[tbp]
\centering 
\includegraphics[width=0.49\textwidth,clip]{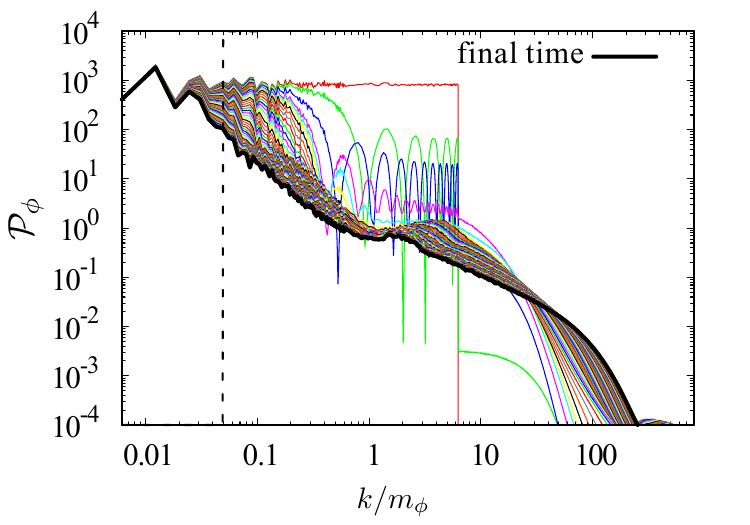}
\includegraphics[width=0.49\textwidth,clip]{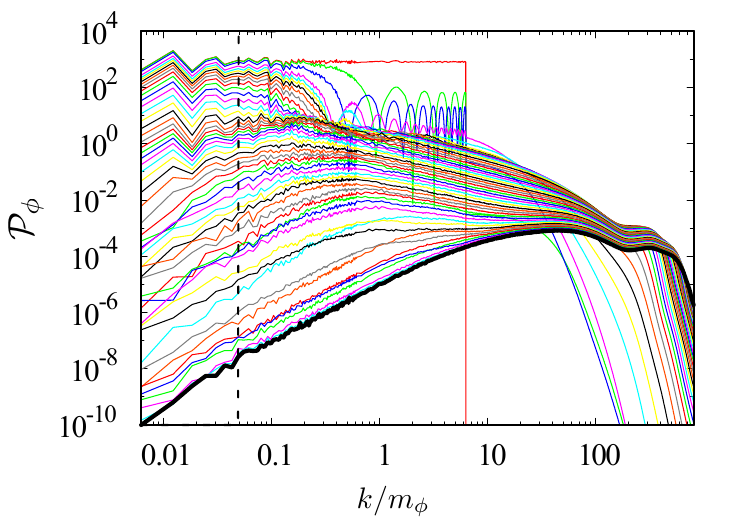}
\caption{Evolution of the power spectrum of the scalar field value for $\tilde{\epsilon} = 0$ (left) and $0.001$ (right).
The vertical dashed line corresponds to $k=2\pi aH$ at the final time.
}
\label{fig:field_power_spectrum}
\end{figure}

In Fig.~\ref{fig:field_power_spectrum}, we show the time evolution of the power spectrum of the axion field with and without the potential bias. Without the potential bias, the spectrum at the final time exhibits a peak at a scale slightly smaller than the Hubble scale. This peak reflects the remaining domain-wall structure. In the presence of the potential bias, on the other hand, the domain walls disappear during the evolution, and the power on the low-$k$ side is significantly suppressed. The remaining peak at higher $k$ reflects axions produced during domain-wall annihilation.

\begin{figure}[tbp]
\centering 
\includegraphics[width=0.49\textwidth,clip]{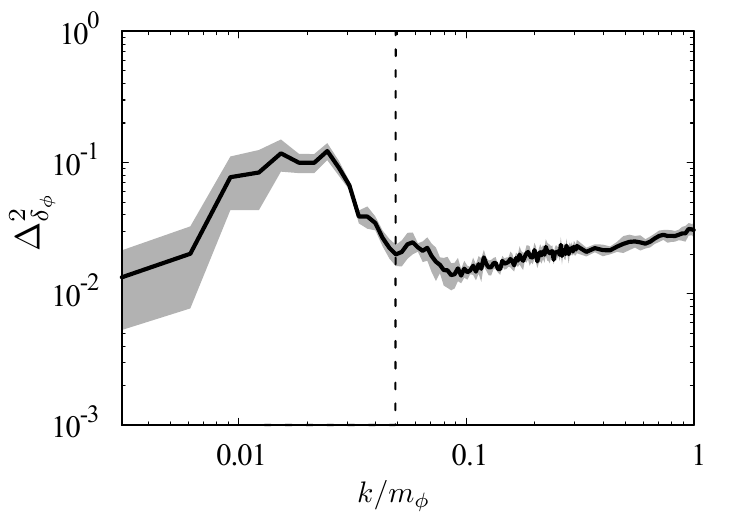}
\includegraphics[width=0.49\textwidth,clip]{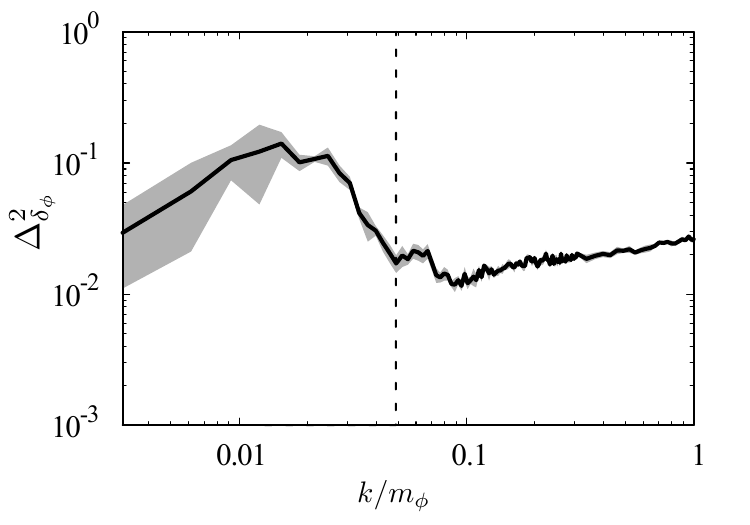}
\includegraphics[width=0.49\textwidth,clip]{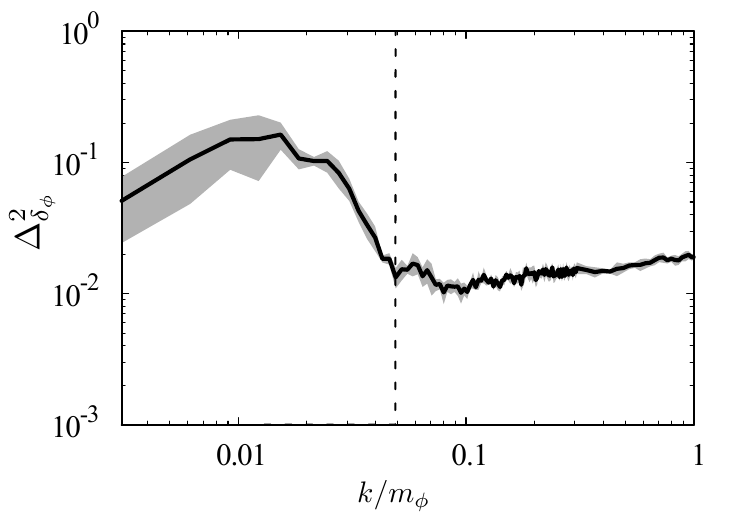}
\includegraphics[width=0.49\textwidth,clip]{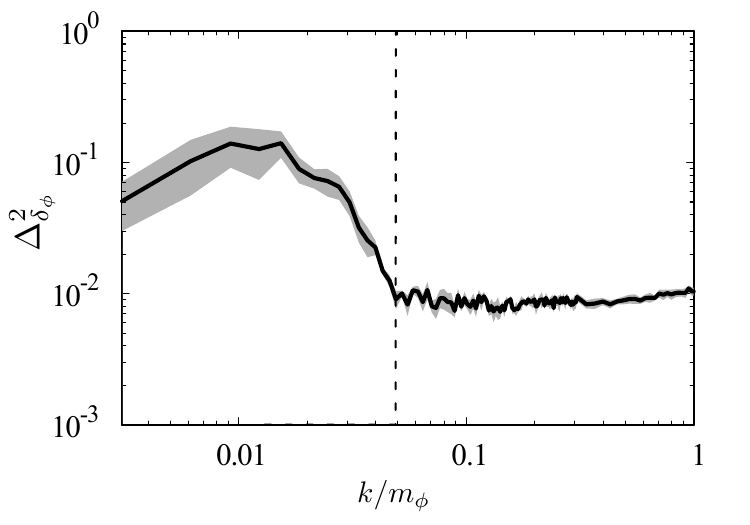}
\caption{The power spectrum of the density contrast of the scalar field at the final time of the simulation ($\tau = 129$) for $\tilde{\epsilon} =0.001$. For the mean and standard deviation for the initial distribution, we have taken
$\langle \phi \rangle = 0$ (top-left),
$\langle \phi \rangle = 6 \pi f_\phi$ (top-right), $12 \pi f_\phi$ (top-left and bottom-left), $24 \pi f_\phi$ (bottom-right) and $\Delta_\phi = 24 \pi f_\phi/\sqrt{\ln(m_\phi L)}$ for all cases.
The vertical dashed line corresponds to the Hubble radius $k = 2 \pi aH$ at the final time, $\tau = 129$.
}
\label{fig:axion_density_power_spectrum2}
\end{figure}

In Fig.~\ref{fig:axion_density_power_spectrum2}, we show the power spectrum of the axion energy-density fluctuations, $\Delta_{\delta_\phi}$, at the final time of the simulations $\tau = 129$. A pronounced peak appears at a wavenumber around the Hubble scale, $k \sim aH$. This feature can be understood as a remnant of the domain-wall collapse. During annihilation, the walls typically sweep over distances of order the Hubble radius, converting the wall and bias energies into axions and thereby generating characteristic density structures on approximately the same scale. The peak is therefore naturally associated with the network dynamics at annihilation.

\begin{figure}[tbp]
\centering 
\includegraphics[width=0.65\textwidth,clip]{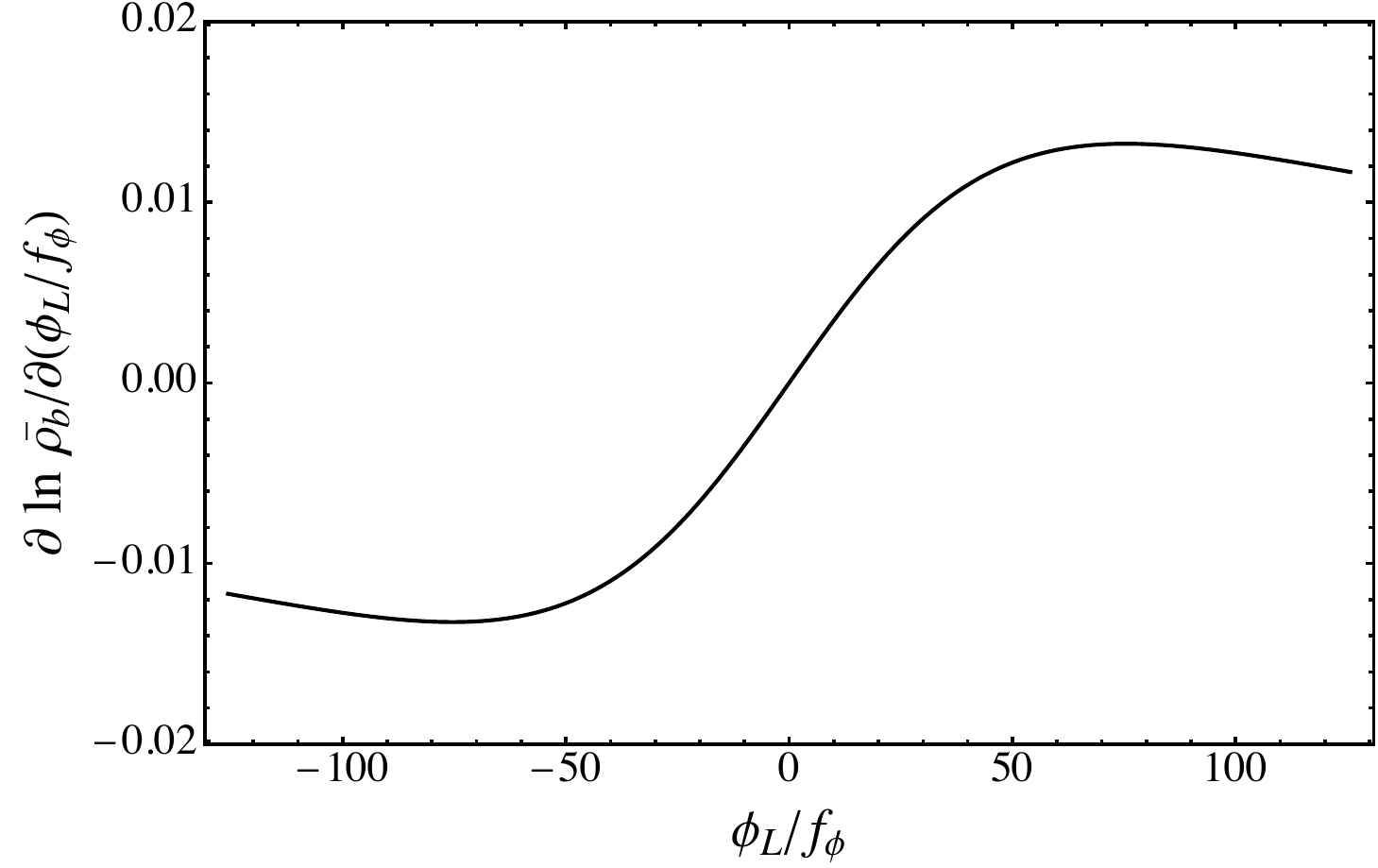}
\caption{The linear response function corresponding to the setup in Fig.~\ref{fig:axion_density_power_spectrum2}. It monotonically increases as $\langle \phi \rangle$ increases from $0$ to $24 \pi f_\phi$.
}
\label{fig:linear_response_2}
\end{figure}

More importantly for the large-scale isocurvature perturbation, the amplitude of the low-$k$ tail increases as $\langle\phi\rangle$ is increased. As can be seen from the linear response function in Fig.~\ref{fig:linear_response_2}, a larger displacement of the mean field generally leads to a stronger modulation of the released bias energy by the long-wavelength field, and hence to a larger nearly scale-invariant contribution to the axion density fluctuations. For $\langle\phi\rangle=0$, the linear response vanishes by symmetry, and the large-scale contribution is expected to decrease %toward zero
at sufficiently small $k$.\footnote{Even when the linear response vanishes, the quadratic response remains nonzero and can account for the nonvanishing axion density power at small $k$.}
For nonzero $\langle\phi\rangle$, by contrast, the response is nonvanishing and the low-$k$ spectrum is expected to approach a scale-invariant plateau whose amplitude increases with $\langle\phi\rangle$ for the parameter range considered here. The present simulations do not extend to sufficiently small $k$ to clearly resolve this asymptotic plateau, but the observed dependence of the low-$k$ amplitude on $\langle\phi\rangle$ is consistent with this expectation.
In fact, after subtracting the power spectrum for the case with $\langle \phi \rangle = 0$, the residual power spectrum is consistent with our analytical estimate for $r_{\rm b}^2 = 0.3$–$0.6$.

%%%%%%%%%%%%%%%%%%%%%%%%%%%%%%%%%%%%%%
\section{Generic Isocurvature constraints on the dark matter from domain wall collapse}
\label{sec:isocurvature-formation}
%%%%%%%%%%%%%%%%%%%%%%%%%%%%%%%%%%%%%%

In the previous section, our lattice simulations demonstrated that superhorizon fluctuations in the bias energy are transferred to the axion energy density after the collapse of the domain wall network.
Motivated by this result,
let us study the isocurvature constraints from the domain wall network in a generic way.\footnote{See also Ref.~\cite{Amin:2022nlh} for the limit on dark matter mass from stochastic isocurvature.} 
Since the dark matter is produced from the collapse of the domain wall network, we consider dark matter whose abundance is determined at a temperature $T_f$.
Here and below, the subscript $f$ refers to the epoch at which dark matter is produced by the network collapse, rather than to the earlier epoch at which the domain-wall network forms.

The dark matter is assumed to behave as pressureless matter, with its comoving abundance remaining constant thereafter.
The comoving wavenumber corresponding to the Hubble scale at the formation epoch is denoted by
\begin{equation}
  k_f \equiv a_f H_f .
\end{equation}
During radiation domination, this scale is approximately
\begin{equation}
  k_f \simeq
  1.7 \times 10^7 \, {\rm Mpc}^{-1}
  \left( \frac{T_f}{1\,{\rm GeV}} \right)
  \left( \frac{g_*(T_f)}{100} \right)^{1/2}
  \left( \frac{g_{*s}(T_f)}{100} \right)^{-1/3},
  \label{eq:kf-radiation}
\end{equation}
where we set the present value of the scale factor to be unity.

We parametrize the large-scale dark matter isocurvature spectrum 
at the formation time by
\begin{equation}
  \Delta_S^2(k)
  =
  A_f
  \left( \frac{k}{k_f} \right)^\alpha ,
  \qquad
  k \lesssim k_f.
  \label{eq:isocurvature-spectrum}
\end{equation}
Here $A_f$ is the dimensionless isocurvature power on the Hubble scale at formation,
$
  A_f \equiv \Delta_S^2(k_f),
$
and $\alpha \, (\geq 0)$ controls the large-scale tail. 
In the domain wall collapse case, for the parameter set adopted in our numerical simulations, we find
\begin{equation}
  A_f \approx {\cal O}(0.01),
\end{equation}
whose order of magnitude is consistent with both the numerical results and the analytic estimate.
As we have shown, $\alpha \approx 0$ for the domain wall from inflationary fluctuations, while the case $\alpha=3$ corresponds to a white-noise spectrum. 
Here we take $\alpha$ arbitrary so that our limit is more general. 
Later, we present a mechanism that generates such a $k$ dependence.

The isocurvature perturbation is a gauge-invariant measure of the relative entropy perturbation between different cosmological components. 
For the total cold dark matter (CDM) component, we define
\begin{equation}
  S_c \equiv \delta_c - \frac{3}{4}\delta_\gamma .
\end{equation}
If only a fraction $f_{\rm DM}$ of the present dark matter is produced by the collapse of the domain-wall network, while the other components carry purely adiabatic fluctuations, the total CDM isocurvature perturbation is suppressed by a factor of $f_{\rm DM}$. Accordingly, its power spectrum is suppressed by $f_{\rm DM}^2$. 
We will include this factor explicitly below. Once dark matter production has ceased and there is no subsequent energy exchange with other components, the superhorizon isocurvature perturbation is conserved. The spectrum generated at the network collapse can therefore be directly compared with observational constraints at the relevant wavenumber.

The CMB anisotropy bound can be written in the approximate form \cite{Planck:2018vyg}
\begin{equation}
  \Delta_S^2(k_*) < \beta_{\rm iso} A_s ,
  \label{eq:cmb-isocurvature-bound}
\end{equation}
where $k_* = 0.05\,{\rm Mpc}^{-1}$ is the pivot scale, $A_s \simeq 2.1 \times 10^{-9}$ is the curvature power spectrum amplitude, and $\beta_{\rm iso}$ is the allowed isocurvature fraction. 
For numerical estimates we use $\beta_{\rm iso}=0.04$, corresponding to a representative value
of the current Planck-level constraint on uncorrelated CDM isocurvature perturbations~\cite{Planck:2018vyg,Planck:2018jri}. 
Combining Eqs.\,\eqref{eq:isocurvature-spectrum} and \eqref{eq:cmb-isocurvature-bound}, and rescaling with the fraction of the axion $f_{\rm DM}$, we obtain
\begin{equation}
  f_{\rm DM}^2 A_f
  \left( \frac{k_*}{k_f} \right)^\alpha
  <
  \beta_{\rm iso} A_s .
  \label{eq:cmb-bound-general-alpha}
\end{equation}
If $\alpha = 0$, this gives an upper bound on $A_f$ as
\begin{align}
    A_f \lesssim 8.4 \times 10^{-11} f_\mathrm{DM}^{-2}
    \ ,
\end{align}
or an upper bound on $f_\mathrm{DM}$ as
\begin{align}
    f_\mathrm{DM} \lesssim 9.2 \times 10^{-5} \left( \frac{A_f}{10^{-2}} \right)^{-1/2}
    \ .
\end{align}
On the other hand, if $\alpha \neq 0$, this limit depends on $k_f$ or equivalently $T_f$.
In particular, when we fix $A_f$ and $f_\mathrm{DM}$, we obtain the lower bound on $T_f$.
Using Eq.\,\eqref{eq:kf-radiation}, the corresponding lower bound on the formation temperature is
\begin{equation}
  T_f \gtrsim
  3\,{\rm keV}\frac{k_*}{0.05\,{\rm Mpc}^{-1}}
  \frac{\left( \frac{f_{\rm DM}^2 A_f}{\beta_{\rm iso} A_s} \right)^{1/\alpha}}{10^3}
  \left( \frac{g_*(T_f)}{100} \right)^{-1/2}
  \left( \frac{g_{*s}(T_f)}{100} \right)^{1/3}.
  \label{eq:Tf-bound-CMB}
\end{equation}
For $\alpha=3$, $A_f = 0.01$, and $f_{\rm DM}=1$, this gives a lower bound of roughly $1$–$2\,\mathrm{keV}$, comparable to small-scale-structure constraints from the Lyman-$\alpha$ forest and Milky Way subhalo counts that are required for successful CDM-like structure formation.

We also consider the prospective sensitivity of small-scale probes. For a probe sensitive primarily to a characteristic wavenumber $k_{\rm p}$, let $\Delta_{\rm p}$ denote the projected upper limit on the dimensionless isocurvature power. The resulting constraint is
\begin{equation}
  f_{\rm DM}^2 A_f
  \left( \frac{k_{\rm p}}{k_f} \right)^\alpha
  <
  \Delta_{\rm p}.
  \label{eq:single-scale-reach}
\end{equation}
The corresponding lower bound on the formation temperature is
\begin{equation}
  T_f \gtrsim
  60\, {\rm eV}\frac{k_{\rm p}}{1\,{\rm Mpc}^{-1}}
  \left( \frac{f_{\rm DM}^2 A_f}{\Delta_{\rm p}} \right)^{1/\alpha}
  \left( \frac{g_*(T_f)}{100} \right)^{-1/2}
  \left( \frac{g_{*s}(T_f)}{100} \right)^{1/3}.
  \label{eq:Tf-bound-single-scale}
\end{equation}

For CMB spectral distortions, the physically relevant observable is not the power at a single wavenumber, but the acoustic energy dissipated and integrated over a finite range of scales. 
We model this finite bandwidth by a top-hat window in $\ln k$,
\begin{equation}
  \left< \Delta_S^2 \right>_{\rm SD}
  \equiv
  \frac{1}{\ln(k_{\rm max}/k_{\rm min})}
  \int_{k_{\rm min}}^{k_{\rm max}} d\ln k \,
  \Delta_S^2(k) .
  \label{eq:SD-window-definition}
\end{equation}
For the power-law spectrum in Eq.\,\eqref{eq:isocurvature-spectrum}, this becomes
\begin{equation}
  \left< \Delta_S^2 \right>_{\rm SD}
  =
  f_{\rm DM}^2 A_f
  \left( \frac{k_{\rm eff}(\alpha)}{k_f} \right)^\alpha ,
  \label{eq:SD-window-power-law}
\end{equation}
where
\begin{equation}
  k_{\rm eff}(\alpha)
  =
  \left[
  \frac{k_{\rm max}^{\alpha}-k_{\rm min}^{\alpha}}
       {\alpha \ln(k_{\rm max}/k_{\rm min})}
  \right]^{1/\alpha}.
  \label{eq:keff-alpha}
\end{equation}
In the numerical plots we take $k_{\rm min}=50\,{\rm Mpc}^{-1}$ and $k_{\rm max}=10^4\,{\rm Mpc}^{-1}$ as an approximate spectral-distortion window. 
We use the illustrative reach
\begin{equation}
  \left< \Delta_S^2 \right>_{\rm SD} < 10^{-2},
  \label{eq:SD-reach}
\end{equation}
motivated by the fact that spectral distortions are sensitive to isocurvature perturbations on scales $1\,{\rm Mpc}^{-1} \lesssim k \lesssim {\rm few}\times 10^4\,{\rm Mpc}^{-1}$ \cite{Chluba:2013dna}. 
A more precise treatment requires replacing the top-hat approximation in Eq.\,\eqref{eq:SD-window-definition} with the appropriate spectral-distortion window function.

Finally, we also consider the prospective sensitivity of 21 cm observations.
The 21 cm signal can be particularly sensitive to blue isocurvature spectra, although the mapping from the primordial isocurvature spectrum to an observable signal depends on the redshift window, astrophysical modeling, and the survey strategy \cite{Takeuchi:2013hza}. 
Therefore, in the figure we do not treat the 21 cm curve as a full forecast. 
Instead, we show an effective single-scale sensitivity,
\begin{equation}
  \Delta_S^2(k_{21}) < \Delta_{21},
  \qquad
  k_{21}=10^2\,{\rm Mpc}^{-1},
  \qquad
  \Delta_{21}=10^{-1}.
  \label{eq:21cm-effective-reach}
\end{equation}
This line is intended only to illustrate the scale at which a 21 cm probe could become relevant. 
A quantitative 21 cm forecast should instead use the corresponding window function or a survey-specific likelihood.

\begin{figure}[t]
  \centering
  \includegraphics[width=0.72\linewidth]{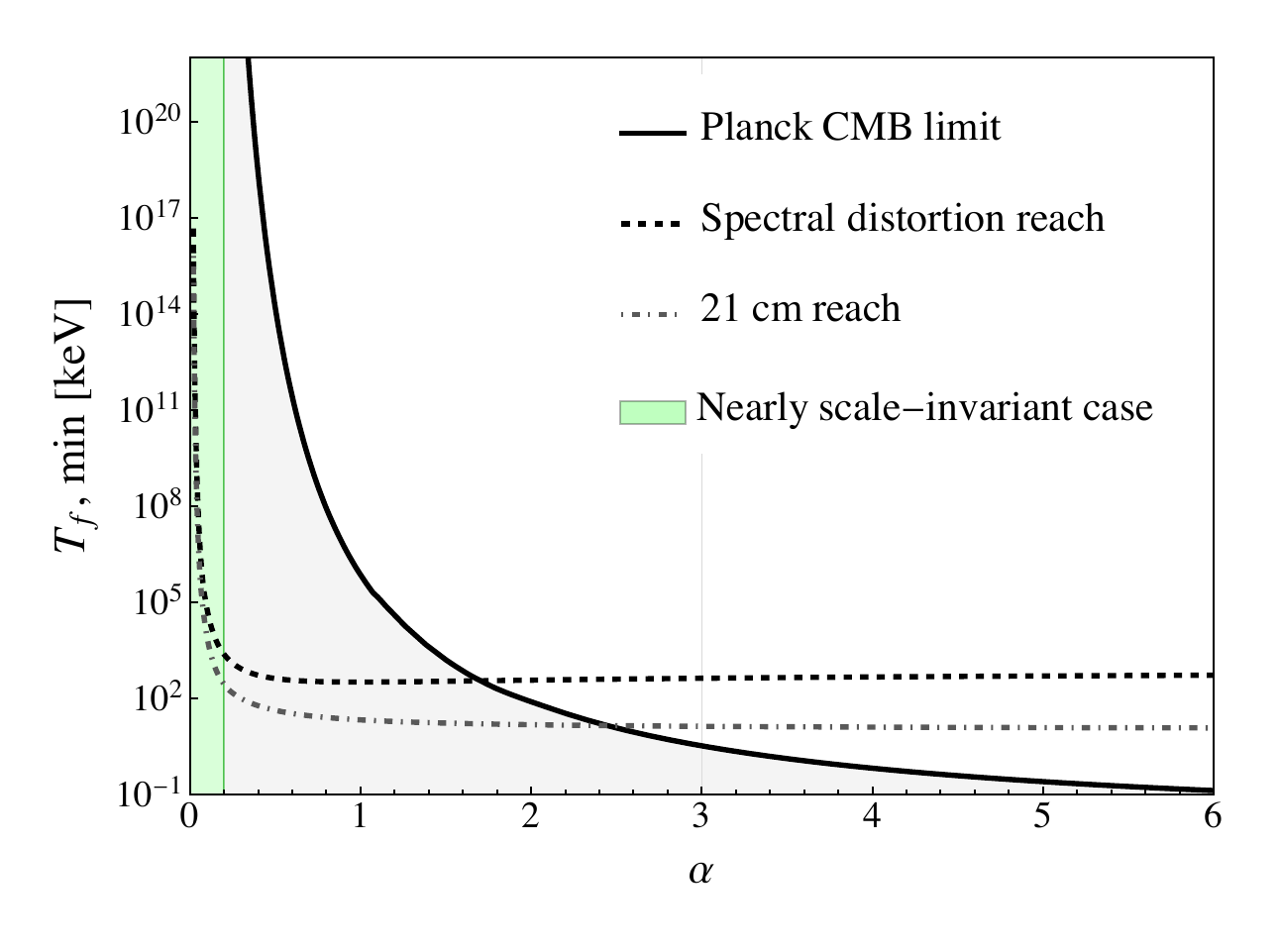}
  \caption{
    Lower bound on the dark matter formation temperature $T_f$ as a function of the isocurvature power index $\alpha$.
    We take $A_f=\Delta_S^2(k_f)=0.02$ and $f_{\rm DM}=1$.
    The shaded region with the solid black curve is excluded by the Planck CMB isocurvature limit.
    The dashed curve shows an effective spectral-distortion reach. 
    The dot-dashed curve shows an indicative 21 cm reach. 
    The green shaded band at $0\leq \alpha \leq 0.2$ indicates the nearly scale-invariant domain-wall fluctuation regime.
  }
  \label{fig:Tf-alpha}
\end{figure}

The resulting constraints are displayed in Fig.~\ref{fig:Tf-alpha}. 
The CMB bound dominates for small $\alpha$, because the power leaks efficiently to the large CMB scales. 
For larger $\alpha$, the CMB constraint weakens rapidly, while small-scale probes become more relevant because they access modes closer to the formation Hubble scale. 
In this sense, the power index $\alpha$ controls how strongly the isocurvature generated at the formation scale is transmitted to observable large or intermediate scales.

We found that dark matter produced by domain-wall collapse from scale-invariant initial fluctuations is excluded by isocurvature constraints, even when the large-scale spectrum is allowed to deviate from exact scale invariance with $\alpha$ as large as $0.3-0.4$. The reason is that the required dark matter formation temperature exceeds $10^{16}\,{\rm GeV}$ and is therefore higher than the maximum temperature attainable after inflation, given the upper bound on the inflationary energy scale inferred from the tensor-to-scalar ratio.

%%%%%%%%%%%%%%%%%%%%%%%%%%%%%%%%%%%%%%
\section{Discussion and Conclusions}
\label{sec: conclusion}
%%%%%%%%%%%%%%%%%%%%%%%%%%%%%%%%%%%%%%

One way to evade the isocurvature constraint on axion dark matter produced from domain walls seeded by inflationary fluctuations is to make the axion decay constant time dependent during inflation~\cite{Linde:1990yj,Linde:1991km,Kasuya:2009up,Kobayashi:2016qld}. 
This can be realized, for example, by coupling the PQ Higgs field to an operator in the inflaton sector,
\begin{equation}
  {\cal L} \supset - |\Phi_{\rm PQ}|^2 {\cal O}_{\rm inf}.
\end{equation}
Since the inflaton background slowly rolls, this interaction can induce a time-dependent expectation value of the PQ Higgs field during inflation. 
As a result, the axion decay constant evolves during inflation, so that different modes experience different effective decay constants at horizon exit.
The angular fluctuation is then estimated as
\begin{equation}
  \delta \theta(k)
  \simeq
  \frac{\delta \phi(k)}{f_\phi(N_k)}
  \simeq
  \frac{H_{\rm inf}(N_k)}{2\pi f_\phi(N_k)},
\end{equation}
where $N_k$ denotes the e-folding time at which the mode $k$ exits the horizon. 
Thus the resulting reduced power spectrum of the angular fluctuation is
\begin{equation}
  \Delta_\theta^2(k)
  \equiv
  \frac{k^3}{2\pi^2} P_\theta(k)
  \simeq
  \left[
    \frac{H_{\rm inf}(N_k)}
         {2\pi f_\phi(N_k)}
  \right]^2 .
  \label{eq:theta-reduced-power}
\end{equation}
Equivalently, normalizing the spectrum at a reference scale $k_*$, we obtain
\begin{equation}
  \frac{\Delta_\theta^2(k)}{\Delta_\theta^2(k_*)}
  =
  \left[
    \frac{H_{\rm inf}(N_k)}{H_{\rm inf}(N_*)}
  \right]^2
  \left[
    \frac{f_\phi(N_*)}{f_\phi(N_k)}
  \right]^2 .
  \label{eq:theta-reduced-power-ratio}
\end{equation}
If $N$ is defined as the number of e-folds before the end of inflation, the horizon-exit condition gives $d\ln k \simeq -dN$. 
Therefore the spectral tilt of the angular fluctuation is
\begin{equation}
  n_\theta - 1
  \equiv
  \frac{d\ln \Delta_\theta^2}{d\ln k}
  \simeq
  -2\epsilon_H
  +
  2\frac{d\ln f_\phi}{dN},
  \label{eq:theta-tilt}
\end{equation}
where $\epsilon_H \equiv -\dot H_{\rm inf}/H_{\rm inf}^2$. 
Thus, if $f_\phi$ was larger at earlier times, the angular fluctuation on large scales is suppressed and the spectrum becomes blue tilted. 
For example, if
\begin{equation}
  f_\phi(N) = f_{\phi,*} e^{\beta (N-N_*)},
\end{equation}
then
\begin{equation}
  \Delta_\theta^2(k)
  =
  \Delta_\theta^2(k_*)
  \left( \frac{k}{k_*} \right)^{2\beta-2\epsilon_H},
\end{equation}
up to slow-roll corrections. 
This time dependence can reduce the large-scale isocurvature perturbation while keeping sizable angular fluctuations on smaller scales.

Assuming that the time variation of $f_\phi$ is absent around the epoch of domain-wall formation, and neglecting $\epsilon_H$, the large-scale tail is approximately characterized by
\begin{equation}
  \alpha \simeq 2\beta .
\end{equation}
Therefore the isocurvature bound discussed above may be evaded for
\begin{equation}
  \beta \gtrsim 0.2 ,
\end{equation}
corresponding to a sufficiently blue angular spectrum, depending on the formation temperature (see Fig.~\ref{fig:Tf-alpha}).
For $\alpha \gtrsim 2$, future improvements in small-scale isocurvature probes may test this scenario. 
Notice also that, in this case, a potential bias may not be necessary: the population bias generated by the inflationary fluctuation itself can make the domain-wall network collapse because it is no longer scale invariant (c.f. \cite{Gonzalez:2022mcx,Diego-Dthesis}).

So far, we have focused on the case of large initial fluctuations in the presence of a periodic potential. We now turn to a simple $\mathbb{Z}_2$ domain-wall model with large initial fluctuations,
\begin{align}
V(\phi)
=
-\frac{m_\phi^2}{2}\phi^2
+\frac{\lambda}{4}\phi^4.
\end{align}
A related scenario was discussed in Ref.~\cite{Kitajima:2023kzu} in the large-volume limit, where the integrated inflationary fluctuation $\langle\delta\phi^2\rangle$ grows with the volume. For
$\langle\delta\phi^2\rangle\gg m_\phi^2/\lambda$,
the fluctuations generate a positive effective mass,
\begin{align}
m_{\rm eff}^2
\simeq
-m_\phi^2
+
3\lambda\langle\delta\phi^2\rangle,
\end{align}
and may temporarily restore the symmetry statistically after the onset of oscillations. The scalar field oscillates around the origin in many Hubble patches and generates subhorizon waves with a characteristic length of order $m_{\rm eff}^{-1}$. As these fluctuations redshift and $m_{\rm eff}$ decreases, the symmetry is broken again, producing domain walls or other topological defects associated with the symmetry of $\phi$.

Our numerical results indicate that scale-invariant fluctuations retain long-range correlations, leaving domain walls and large void-like regions after the quartic-induced waves have sufficiently redshifted. Even in the presence of a population bias, large initial fluctuations do not necessarily eliminate the walls. Instead, the fluctuation-induced effective mass damps the fluctuations before the transition and can make the resulting network more persistent. Thus, symmetry restoration driven by scale-invariant fluctuations does not readily solve the domain-wall problem: large-scale topological defects may survive the subsequent phase transition. A more detailed investigation is left for future work.

In this paper, we have revisited the evolution of axion domain walls seeded by inflationary fluctuations, focusing on the regime in which the initial fluctuations populate many vacua.  Using analytical arguments and numerical lattice simulations, we studied both the subsequent wall dynamics and the isocurvature perturbations carried by axions produced during domain-wall annihilation.

Our main result is that populating many vacua does not generically erase the primordial isocurvature perturbations.  Although the conventional misalignment contribution can be smoothed by averaging over many vacua, the energy released during domain-wall annihilation depends on the long-wavelength axion field and therefore inherits its superhorizon correlations.  The resulting axion density perturbations can thus remain sizable even after the wall network has entered the scaling regime and disappeared.  This suggests that evading axion isocurvature constraints solely by populating multiple vacua is generically difficult.

We also discussed the formation of composite domain walls when multiple periods of the full potential are populated, as well as isocurvature constraints for a general power-law spectrum with a nonstandard spectral index.  A more detailed study of gravitational-wave production in the multiple-vacuum regime, including the possible long-lived evolution of composite walls, is left for future work.

While we have mainly focused on the regime in which the populated field range covers only a single period of the bias potential, the multiple-bias-period regime may also lead to characteristic observational signatures.
In particular, for $f_{\rm b} \lesssim H_{\rm inf} \lesssim F$, the domain-wall network eventually disappears, but its collapse can proceed over an extended period.
Elementary walls separating adjacent local vacua with relatively large energy differences disappear first, whereas the resulting composite walls survive until a later epoch.
Therefore, gravitational waves produced during these two stages may have two peaks at different frequencies.
Moreover, since the elementary walls collapse sequentially rather than at a single well-defined time, the corresponding peak will become broader than in the conventional domain wall collapse.
A detailed investigation of the resulting gravitational-wave spectrum is an interesting direction for future work.

%%%%%%%%%%%%%%%%%%%%%%%%%%%%%%%%%%%%%
\section*{Acknowledgments}
%%%%%%%%%%%%%%%%%%%%%%%%%%%%%%%%%%%%%
This work was supported by JSPS KAKENHI Grant Numbers 22K14029(W.Y.), 23K22486 (W.Y.), 24K17039(K.M.), 25H02165 (F.T.), 25KJ0564 (J.L.), and 26K00695 (F.T., N.K., and W.Y.).
This work was also supported by the World Premier International Research Center Initiative (WPI), MEXT, Japan, and by COST Action COSMIC WISPers CA21106, supported by COST (European Cooperation in Science and Technology). 
J.L. is also supported by the Graduate Program on Physics for the Universe (GP-PU), Tohoku University.
W.Y. is also supported by the Selective Research Fund and the Incentive Research Fund of Tokyo Metropolitan University.
This work used computational resources of Fugaku supercomputer, provided by RIKEN Center for Computational Sciences, through the HPCI System Research Project (Project ID: hp260117). 

%%%%%%%%%%%%%%% References %%%%%%%%%%%%%%%%
\bibliographystyle{apsrev4-1}
\bibliography{ref}
%%%%%%%%%%%%%%%%%%%%%%%%%%%%%%%%%%%%%%%%%%%

\end{document}